\documentclass[twocolumn,showpacs,superscriptaddress,pre,nofootinbib]{revtex4}

\usepackage{graphicx}
\usepackage{amsmath,amssymb}
\usepackage{color}
\usepackage{amsfonts}

\usepackage{subfig}

\begin{document}

\newtheorem{theorem}{Theorem}
\newtheorem{definition}{Definition}
\newtheorem{lemma}{Lemma}
\newtheorem{proposition}{Proposition}
\newtheorem{remark}{Remark}
\newtheorem{corollary}{Corollary}
\newtheorem{example}{Example}


\title{Stochastic resetting on comb-like structures\footnote{V. D. and A. M. contributed equally to this work.}}


\author{Viktor Domazetoski}
\email{v.domazetoski@manu.edu.mk}
\affiliation{Research Center for Computer Science and Information Technologies, Macedonian Academy of Sciences and Arts, Bul. Krste Misirkov 2, 1000 Skopje, Macedonia}

\author{Axel Mas\'o-Puigdellosas}
\email{Axel.Maso@uab.cat}
\affiliation{Grup de F\'isica Estad\'istica, Departament de F\'isica. Universitat Aut\`onoma de Barcelona. Edifici Cc. E-08193 Cerdanyola (Bellaterra) Spain}

\author{Trifce Sandev}
\email{trifce.sandev@manu.edu.mk} \affiliation{Research Center for Computer Science and Information Technologies, Macedonian Academy of Sciences and Arts, Bul. Krste Misirkov 2, 1000 Skopje, Macedonia} \affiliation{Institute of Physics \& Astronomy, University of Potsdam, D-14776 Potsdam-Golm, Germany} \affiliation{Institute of Physics, Faculty of Natural Sciences and Mathematics, Ss.~Cyril and Methodius University, Arhimedova 3, 1000 Skopje, Macedonia}

\author{Vicen{\c{c}} M\'endez}
\email{Vicenc.Mendez@uab.cat}
\affiliation{Grup de F\'isica Estad\'istica, Departament de F\'isica. Universitat Aut\`onoma de Barcelona. Edifici Cc. E-08193 Cerdanyola (Bellaterra) Spain}

\author{Alexander Iomin}
\email{iomin@physics.technion.ac.il}
\affiliation{Department of Physics, Technion, Haifa 32000, Israel}

\author{Ljupco Kocarev}
\email{lkocarev@manu.edu.mk}
\affiliation{Research Center for Computer Science and Information Technologies, Macedonian Academy of Sciences and Arts, Bul. Krste Misirkov 2, 1000 Skopje, Macedonia}
\affiliation{Faculty of Computer Science and Engineering, Ss.~Cyril and Methodius University,\\ P.O. Box 393, 1000 Skopje, Macedonia}

\date{Submitted 12 Dec 2019, Revised 25 Apr 2020} 

\begin{abstract}
We study a diffusion process on a three-dimensional comb under stochastic resetting. We consider three different types of resetting: global resetting from any point in the comb to the initial position, resetting from a finger to the corresponding backbone and resetting from secondary fingers to the main fingers. The transient dynamics along the backbone in all three cases is different due to the different resetting mechanisms, finding a wide range of dynamics for the mean squared displacement. For the particular geometry studied herein, we compute the stationary solution and the mean square displacement and find that the global resetting breaks the transport in the three directions. Regarding the resetting to the backbone, the transport is broken in two directions but it is enhanced in the main axis. Finally, the resetting to the fingers enhances the transport in the backbone and the main fingers but reaches a steady value for the mean squared displacement in the secondary fingers.
\end{abstract}

\maketitle

\section{Introduction}
Combs are two- or three-dimensional branched structures with a backbone crossed by perpendicular fingers. These fingers may be one or two dimensional side structures. A random walker moving along the backbone may enter into a finger (or fingers) and move there for a time and return to the backbone to start the process again. As a result of a Brownian motion in a two-dimensional comb the mean squared displacement (MSD) shows a subdiffusive behavior depending on time as $t^{1/2}$ and was originally introduced to understand anomalous transport in percolation clusters and many other applications \cite{book}. Although three-dimensional combs have been less developed, they have modeled transport in spiny dendrites \cite{MeIo13} or ultraslow diffusion in combs with circular fingers \cite{IoMe16}.

\begin{figure}
\centering{\includegraphics[width=8cm]{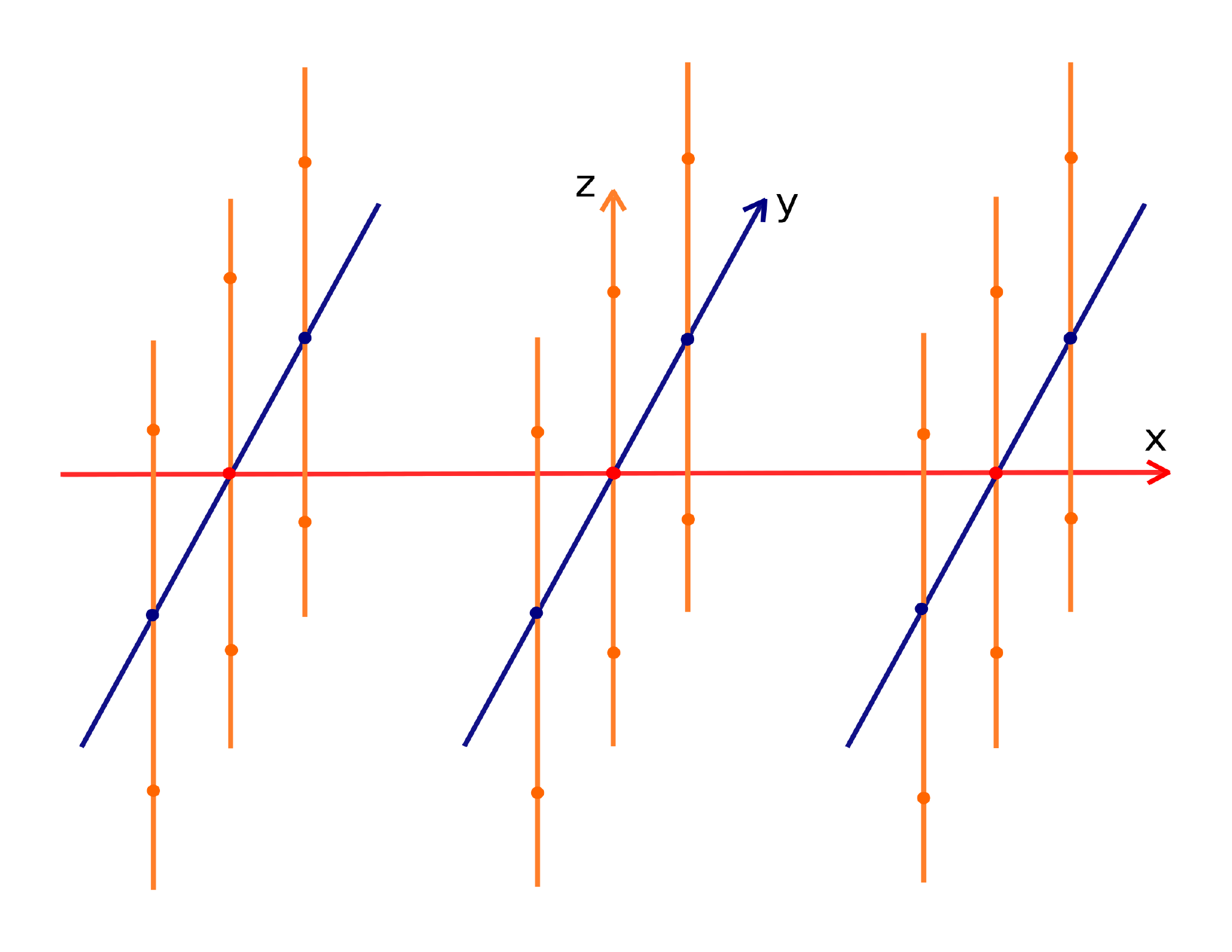}
\caption{Three-dimensional comb-like structure,
which is a discrete caricature of the continuous 3D comb model 
described by Eq. \eqref{diffusion eq 3Dcomb}. 
It consists of the backbone along the $x$ axis and continuously distributed
side-branches - fingers along the $y$ and $z$ axes. \label{fig_comb}}
\label{Fig1}}
\end{figure}

In this paper we investigate a diffusion process on a $xyz$-comb (see Fig.~\ref{fig_comb}) with stochastic resetting. The diffusion on a three dimensional comb is governed by the following equation
\begin{align}\label{diffusion eq 3Dcomb}
\frac{\partial}{\partial t}P(x,y,z,t)=L_{\mathrm{FP}}P(x,y,t), 
\end{align}
where $$L_{\mathrm{FP}}=\mathcal{D}_{x}\delta(y)\delta(z)\frac{\partial^2}{\partial x^2}+\mathcal{D}_{y}\delta(z)\frac{\partial^2}{\partial y^2}+\mathcal{D}_{z}\frac{\partial^2}{\partial z^2}$$ is the Fokker-Planck (transport) operator, and $\mathcal{D}_{x}\delta(y)\delta(z)$, $\mathcal{D}_{y}\delta(z)$ and $\mathcal{D}_{z}$ are the diffusion coefficients along the $x$, $y$ and $z$ directions, respectively. The $\delta$-functions $\delta(y)\delta(z)$ in front of the second spatial derivative with respect to $x$, mean that diffusion along the backbone ($x$-axis) is allowed only at $y=z=0$, while the $\delta$-function $\delta(z)$ in front of the second spatial derivative with respect to $y$, means that the diffusion along the main fingers (or branches) ($y$-axis) is allowed only at $z=0$. The $z$-axis is a secondary finger, an auxiliary direction along which the particle performs normal diffusion.

On the other hand, the diffusion process in one dimension under stochastic resetting was introduced by Evans and Majumdar \cite{maj1}. The corresponding equation is given by
{\small{\begin{align}\label{diffusion eq resetting sol}
\frac{\partial}{\partial t}P(x,t|x_0)=\mathcal{D}\frac{\partial^2}{\partial x^2}P(x,t|x_0)-rP(x,t|x_0)+r\delta(x-x_0),
\end{align}
}}where the initial position reads $P(x,t=0|x_0)=\delta(x-x_0)$, $\mathcal{D}$ is the diffusion coefficient, $r$ is the rate of resetting to the initial position, the second term on the right hand side represents the loss of probability from the position $x$ due to reset to the initial position $x_0$, and the third term is the gain of probability at $x_0$ due to resetting from all other positions. This equation represents a renewal process: each resetting event to the initial position $x_0$ renews the process at a rate $r$. Between two consecutive renewal events, the particle undergoes free diffusion \cite{maj1}. It is known that this equation has a stationary solution in the long time limit given by $$P_{\mathrm{st}}(x|x_0)=\frac{1}{\sqrt{4\mathcal{D}/r}}e^{-\frac{|x-x_0|}{\sqrt{\mathcal{D}/r}}}.$$

Afterwards, other types of motion and resetting mechanisms have been studied by introducing the two resetting terms to the Fokker-Planck equation of the corresponding process \cite{maj2,EvMa14,maj3,DuHe14,Pa15,ChSc15,Ma19,Gu19,lenzi_front}. For instance, space-dependent reset rates \cite{maj2} or diffusion in a potential landscape \cite{Pa15} and the telegraphers equation \cite{Ma19} have been analyzed under this perspective. Other works have studied motion with resetting by employing a renewal equation \cite{MoVi13,mendez,pal1,NaGu16,EuMe16,MoMaVi17,Sh18,EvMa18,axel,MaCaMe19p,BoChSo19,BoChSo19p,MaCaMe19pp,MaCaMe19ppp}, which has also been used to study the completion time of search processes with resetting \cite{CaMe15,reu,pal reu,chechkin,pal pre, pal prr,masoliver2019}.

Multi-dimensional diffusion has already been studied in the literature \cite{EvMa14}. There, diffusion with resets in an multi-dimensional, homogeneous, infinite media is studied. In this work we analyze the transport properties and the long time behavior of diffusion in an heterogeneous environment and determine the properties emerging from resetting in the different space coordinates.

The paper is organized as follows. In Section II we consider diffusion in a three-dimensional comb with global exponential (Markovian) resetting. We give exact results for the marginal probability density functions (PDFs), stationary distributions and MSDs along all three axes. We also confirm the analytical results by numerical simulations by employing a Langevin equation approach for comb structure. Excellent agreement has been shown. Diffusion in three-dimensional comb with exponential resetting to the backbone is considered in Section III and the corresponding PDFs and MSDs are also found. In Section IV exponential resetting to the fingers is analyzed. We also discuss the resetting mechanisms in two-dimensional comb-structures in Section V. In Section VI we give detailed explanation of the topological constraint of the transport properties of both two- and three-dimensional comb structures. Summary is provided in Section VII.

\section{Global resetting}

\subsection{Analytical results}

We start our analysis by considering diffusion in a three dimensional comb with global resetting, represented by the equation
{\small{\begin{align}\label{diffusion eq 3Dcomb resetting}
&\frac{\partial}{\partial t}P(x,y,z,t|x_0,0,0)=L_{\mathrm{FP}}P(x,y,z,t|x_0,0,0) 
\nonumber\\&-rP(x,y,z,t|x_0,0,0)+r\delta(x-x_0)\delta(y)\delta(z),
\end{align}
}}with the initial position $P(x,y,z,t=0|x_0)=\delta(x-x_0)\delta(y)\delta(z)$. This equation can also be interpreted in terms of a renewal process: each resetting event to the initial position $(x_0,y_0,z_0)=(x_0,0,0)$ renews the process at a rate $r$. Between two consecutive renewal events, the particle undergoes diffusion on the $xyz$-comb structure.

To find the solution of Eq.~(\ref{diffusion eq 3Dcomb resetting}) we apply the Fourier transformations\footnote{The Fourier transform of a function $f(\xi)$ is given by $f(k)=\mathcal{F}[f(x)](k)=\int_{-\infty}^{\infty}f(\xi)\,e^{\imath k \xi}\, d\xi$. The inverse Fourier transform then reads $f(\xi)=\mathcal{F}^{-1}[f(k)](x)=\frac{1}{2\pi}\int_{-\infty}^{\infty}f(k)\,e^{-\imath k \xi}\, dk$.} with respect to $x$, $y$ and $z$, and the Laplace transformation\footnote{The Laplace transform of a function $f(t)$ reads $\hat{f}(s)=\mathcal{L}[f(t)](s)=\int_{0}^{\infty}f(t)\,e^{-st}\,dt$.} with respect to $t$. Therefore, for the PDF in the Fourier-Laplace domain we obtain, see {\it Supplemental material 1} for details of calculations,
\begin{align}\label{P(kx,ky,kz,s) final}
&\hat{P}(k_x,k_y,k_z,s|x_0,0,0)=\frac{1}{s}\times\frac{(s+r)^{1/4}}{(s+r)^{1/4}+\frac{\mathcal{D}_{x}}{2\sqrt{2\mathcal{D}_{y}\sqrt{\mathcal{D}_{z}}}}k_{x}^{2}}\nonumber\\&\times\frac{(s+r)^{1/2}}{(s+r)^{1/2}+\frac{\mathcal{D}_{y}}{2\sqrt{\mathcal{D}_{z}}}k_{y}^{2}}\times\frac{(s+r)}{(s+r)+\mathcal{D}_{z}k_{z}^{2}}\times e^{\imath k_{x}x_{0}}.\nonumber\\
\end{align}

\subsection{Marginal PDFs}

In order to analyze the motion along all three directions, we analyze the marginal PDFs,
{\small{\begin{align}
&p_{1}(x,t|x_0)=\int_{-\infty}^{\infty}\int_{-\infty}^{\infty}P(x,y,z,t|x_0,0,0)\,dy\,dz\label{marginalPDFx}\\&
p_{2}(y,t|0)=\int_{-\infty}^{\infty}\int_{-\infty}^{\infty}P(x,y,z,t|x_0,0,0)\,dx\,dz,\label{marginalPDFy}\\&
p_{3}(z,t|0)=\int_{-\infty}^{\infty}\int_{-\infty}^{\infty}P(x,y,z,t|x_0,0,0)\,dx\,dy.\label{marginalPDFz}
\end{align}
}}In the Fourier-Laplace space, the marginal PDFs are
\begin{align}\label{p1_fl}
    \hat{p}_1(k_x,s|x_0)=\hat{P}(k_x,k_y=0,k_z=0,s|x_0,0,0),
\end{align}
\begin{align}\label{p2_fl}
    \hat{p}_2(k_y,s|0)=\hat{P}(k_x=0,k_y,k_z=0,s|x_0,0,0),
\end{align}
\begin{align}\label{p3_fl}
    \hat{p}_3(k_z,s|0)=\hat{P}(k_x=0,k_y=0,k_z,s|x_0,0,0).
\end{align}

Therefore, from Eqs.~(\ref{P(kx,ky,kz,s) final}) and (\ref{p1_fl}), for the marginal PDF along the backbone we have
\begin{align}\label{p1 global}
\hat{p}_{1}(k_x,s|x_0)=\frac{s^{-1}(s+r)^{1/4}}{(s+r)^{1/4}+\mathcal{D}_{1}\,k_{x}^{2}}\,e^{\imath k_x x_0},
\end{align}
where $\mathcal{D}_{1}=\frac{\mathcal{D}_{x}}{2\sqrt{2\mathcal{D}_{y}\sqrt{\mathcal{D}_{z}}}}$. By applying the inverse Fourier transform we obtain
\begin{align}
\hat{p}_{1}(x,s|x_0)=\frac{1}{2\sqrt{\mathcal{D}_{1}}}s^{-1}(s+r)^{1/8}e^{-\frac{(s+r)^{1/8}}{\sqrt{\mathcal{D}_{1}}}|x-x_0|}.
\end{align}
From Eq.~(\ref{p1 global}), by inverse Fourier-Laplace transforms we arrive at the generalized (non-Markovian) diffusion equation along the backbone
\begin{align}
\int_{0}^{t}\gamma(t-t')\frac{\partial}{\partial t'}p_{1}(x,t'|x_0)\,dt'=\mathcal{D}_{1}\frac{\partial^{2}}{\partial x^2}p_{1}(x,t|x_0),
\end{align}
where
\begin{align}
    \gamma(t)
=t^{1/4}E_{1,5/4}^{1/4}(-r\,t).
\end{align}
Here
\begin{equation}\label{ML three}
E_{\alpha,\beta}^{\delta}(z)=\sum_{k=0}^{\infty}\frac{(\delta)_k}{\Gamma(\alpha k+\beta)}\frac{z^k}{k!}
\end{equation}
is the three parameter Mittag-Leffler function \cite{prabhakar}\footnote{The Laplace transform of the three parameter Mittag-Leffler function reads \cite{prabhakar}
$$\mathcal{L}\left[t^{\beta-1}E_{\alpha,\beta}^{\delta}(\pm at^{\alpha})\right](s)=\frac{s^{\alpha\delta-\beta}}{\left(s^{\alpha}\mp a\right)^{\delta}}, \quad \Re(s)>|a|^{1/\alpha}.$$
Its asymptotic behaviors are given by \cite{garrappa,pre2015}
$$E_{\alpha,\beta}^{\gamma}(-z^{\alpha})\simeq\left\lbrace\begin{array}{ll}
     \frac{1}{\Gamma(\beta)}\exp\left(-\gamma\frac{\Gamma(\beta)}{\Gamma(\alpha+\beta)}z^{\alpha}\right), \quad z\ll1.  \\
     \frac{z^{-\alpha\gamma}}{\Gamma(\beta-\alpha\gamma)}, \quad z\gg1.
\end{array}\right.$$}
and $(\delta)_k=\Gamma(\delta+k)/\Gamma(\delta)$ is the Pochhammer symbol. The initial condition is $p_{1}(x,t=0|x_0)=\delta(x-x_0)$. The equation can also be written in the form
\begin{align}
{_C}\mathbf{D}_{1,-r,0+}^{1/4,1/4}p_{1}(x,t|x_0)=\mathcal{D}_{1}\frac{\partial^2}{\partial x^2}p_{1}(x,t|x_0),
\end{align}
where
\begin{align}\label{prabhakar derivative}
{_C}\mathbf{D}_{\rho,-\nu,0+}^{\delta,\mu}f(t)=\int_{0}^{t}(t-t')^{-\mu}E_{\rho,1-\mu}^{-\delta}\left(-\nu t^{\rho}\right)\frac{df(t')}{dt'}\,dt',
\end{align}
is a so-called regularized Prabhakar derivative \cite{polito}, which has many applications nowadays \cite{sandev mathematics,epl,weron}. The stationary PDF along the backbone, obtained in the long time limit, becomes
\begin{align}
p_{1,\mathrm{st}}(x|x_0)=\frac{1}{\sqrt{4\mathcal{D}_{1}/\sqrt[4]{r}}}e^{-\frac{|x-x_0|}{\sqrt{\mathcal{D}_{1}/\sqrt[4]{r}}}}.
\label{StDistxGlobal}
\end{align}

From Eqs.~(\ref{P(kx,ky,kz,s) final}) and (\ref{p2_fl}), for the PDF along the fingers we find
\begin{align}\label{P(kx,ky,kz,s) final2}
\hat{p}_{2}(k_y,s|0)&=\frac{s^{-1}(s+r)^{1/2}}{(s+r)^{1/2}+\mathcal{D}_{2}\,k_{y}^{2}},
\end{align}
where $\mathcal{D}_{2}=\frac{\mathcal{D}_{y}}{2\sqrt{\mathcal{D}_{z}}}$. From here, by applying the inverse Fourier transform we obtain
\begin{equation}
 \hat{p}_{2}(y,s|0)=\frac{(s+r)^{1/4}}{2s\sqrt{\mathcal{D}_{2}}}e^{-\frac{(s+r)^{1/4}}{\sqrt{\mathcal{D}_{2}}}|y|}.
\label{P(kx,ky,kz,s) final2}
\end{equation}
The marginal PDF $p_2(y,t|0)$ provides the transport equation along the main finger and it is governed by the equation
\begin{align}\label{pdf2_global}
\int_{0}^{t}\zeta(t-t')\frac{\partial}{\partial t'}p_{2}(y,t'|0)\,dt'=\mathcal{D}_{2}\frac{\partial^{2}}{\partial y^2}p_{2}(y,t|0),
\end{align}
with the initial condition $p_2(y,t=0|0)=\delta(y)$. Here the kernel
$\zeta(t)$ is
\begin{align}
\zeta(t)
=t^{-1/2}E_{1,1/2}^{-1/2}(-rt)=\frac{1}{\sqrt{\pi t}}e^{-rt}+\sqrt{r}\,\mathrm{erf}\left(\sqrt{rt}\right).
\end{align}
Eq.~(\ref{pdf2_global}) can also be presented by means of the regularized Prabhakar derivative (\ref{prabhakar derivative}). It reads
\begin{align}
{_C}\mathbf{D}_{1,-r,0+}^{1/2,1/2}p_{2}(y,t|0)=\mathcal{D}_{2}\frac{\partial^{2}}{\partial y^2}p_{2}(y,t|0),
\end{align}
or equivalently
\begin{align}
&{_{TC}}D_{r}^{1/2}p_{2}(y,t|0)=\mathcal{D}_{2}\frac{\partial^{2}}{\partial y^2}p_{2}(y,t|0)\nonumber\\&-\sqrt{r}\int_{0}^{t}\mathrm{erf}\left(\sqrt{r(t-t')}\right)\frac{\partial}{\partial t'}p_{2}(y,t'|0)\,dt',
\end{align}
where $\mathrm{erf}(z)=\frac{2}{\sqrt{\pi}}\int_{0}^{z}e^{-t^{2}}\,dt$ is the error function, while
\begin{align}\label{TC derivative}
{_{TC}}D_{b}^{\alpha}f(t)=\frac{1}{\Gamma(1-\alpha)}\int_{0}^{t}e^{-b(t-t')}(t-t')^{-\alpha}\frac{d}{d t'}f(t')\,dt'
\end{align}
is the tempered Caputo derivative with the exponential truncation, where $b>0$ is the truncation parameter \cite{fcaa2015,sandev mathematics}. For the stationary PDF along the $y$ direction, we find
\begin{equation} p_{2,\mathrm{st}}(y|0)=\frac{1}{\sqrt{4\mathcal{D}_{2}/\sqrt{r}}}e^{-\frac{|y|}{\sqrt{\mathcal{D}_{2}/\sqrt{r}}}}.
\label{StDistyGlobal}
\end{equation}

For the $z$ direction, we have
\begin{align}\label{global p3}
\hat{p}_{3}(k_z,s|0)=\frac{s^{-1}(s+r)}{(s+r)+\mathcal{D}_{z}k_{z}^{2}},
\end{align}
that yields
\begin{align}
\hat{p}_{3}(z,s|0)=\frac{1}{2\sqrt{\mathcal{D}_{z}}}s^{-1}(s+r)^{1/2}e^{-\frac{(s+r)^{1/2}}{\sqrt{\mathcal{D}_{3}}}|z|},
\end{align}
where $\mathcal{D}_{3}=\mathcal{D}_{z}$. The corresponding equation for the transport along secondary fingers reads
\begin{align}
\int_{0}^{t}\xi(t-t')\frac{\partial}{\partial t'}p_{3}(z,t'|0)\,dt'=\mathcal{D}_{3}\frac{\partial^{2}}{\partial z^2}p_{3}(z,t|0).
\end{align}
where
\begin{align}
    \xi(t)
    =\delta(t)+r,
\end{align}
and it can be rewritten in the equivalent form
\begin{align}
\frac{\partial}{\partial t}p_{3}(z,t|0)=\mathcal{D}_{3}\frac{\partial^{2}}{\partial z^2}p_{3}(z,t|0)-rp_{3}(z,t|0),
\end{align}
which is a diffusion-absorption equation. The stationary PDF along the $z$ direction is
\begin{align}
p_{3,\mathrm{st}}(z|0)=\frac{1}{\sqrt{4\mathcal{D}_{3}/r}}e^{-\frac{|z|}{\sqrt{\mathcal{D}_{3}/r}}}.
\label{StDistzGlobal}
\end{align}
It is interesting to note that the obtained stationary PDFs along each axis are exponential functions, so that the global resetting does not modify the stationary state with respect to the case of standard diffusion in one dimension under resetting, see Eq.~(\ref{diffusion eq resetting sol}). The global resetting affects the transient dynamics towards the stationary state only. 
For a given value of $\mathcal{D}$ for all three components, the width of the PDF varies. This can be seen from the analytical formulas, in which the width of the exponential stationary PDF depends on $r$ for the $z$ component, on $\sqrt{r}$ for the $y$ component and on $\sqrt[4]{r}$ for the $x$ component, see Eqs.~\eqref{StDistxGlobal}, \eqref{StDistyGlobal}, \eqref{StDistzGlobal}. The obtained analytical results are verified by  stochastic simulations with the Langevin equation approach, see Section~\ref{num simul}.

\subsection{Mean squared displacements}

In the section, we analyze the MSDs along all three directions,
$$\left\langle x^{2}(t)\right\rangle=\int_{-\infty}^{\infty}x^{2}\,p_{1}(x,t|x_0)\,dx,$$
$$\left\langle y^{2}(t)\right\rangle=\int_{-\infty}^{\infty}y^{2}\,p_{2}(y,t|0)\,dy,$$
$$\left\langle z^{2}(t)\right\rangle=\int_{-\infty}^{\infty}z^{2}\,p_{3}(z,t|0)\,dz.$$
Taking into account corresponding solutions for the marginal PDFs, we have
\begin{align}
    \left\langle x^{2}(t)\right\rangle
    &=x_{0}^{2}+2\,\mathcal{D}_{1}\,t^{1/4}E_{1,5/4}^{1/4}(-rt)\nonumber\\&
    =x_{0}^{2}+2\,\mathcal{D}_{1}\,t^{1/4}\left[\frac{1}{\sqrt[4]{rt}}-\frac{\textrm{E}_{3/4}(rt)}{\Gamma(1/4)}\right],
\end{align}
where $\textrm{E}_{n}(z)=\int_{1}^{\infty}\frac{e^{-zt}}{t^{n}}\,dt$ is the exponential integral function. This corresponds to the transition from
subdiffusion to localization
$$\left\langle x^{2}(t)\right\rangle\sim x_{0}^{2}+2\,\mathcal{D}_{1}\left\lbrace\begin{array}{l l}
     & \frac{t^{1/4}}{\Gamma(5/4)}, \quad rt\ll1\\ \\
     & \frac{1}{\sqrt[4]{r}}, \quad rt\gg1.
\end{array}\right.$$ For the $y$ fingers we have
\begin{align}
    \left\langle y^{2}(t)\right\rangle
    =2\,\mathcal{D}_{2}\,\frac{\mathrm{erf}(\sqrt{rt})}{\sqrt{r}},
\end{align}
that corresponds to the transition from
subdiffusion to localization as well,
$$\left\langle y^{2}(t)\right\rangle\sim2\,\mathcal{D}_{2}\left\lbrace\begin{array}{l l}
     & \frac{t^{1/2}}{\Gamma(3/2)}, \quad rt\ll1,\\ \\
     & \frac{1}{\sqrt{r}}, \quad rt\gg1,
\end{array}\right.$$Eventually, the MSD for $z$- fingers reads
\begin{align}
    \left\langle z^{2}(t)\right\rangle
    =2\,\mathcal{D}_{3}\,\frac{1-e^{-rt}}{r},
\end{align}
that corresponds to saturation in the long time limit,
$$\left\langle z^{2}(t)\right\rangle\sim2\,\mathcal{D}_{3}\left\lbrace\begin{array}{l l}
     & t, \quad rt\ll1 \\ \\
     & \frac{1}{r}, \quad rt\gg1.
\end{array}\right.$$
Therefore, unlike an initial transient behaviour all the MSDs saturate towards a constant value (exhibiting stochastic localization) as in the scenario of one-dimensional diffusion with resets. This confirms the existence of a non-equilibrium stationary state, which has been recently observed for many different dynamics under constant-rate resets \cite{axel}. This variety of cases has been also obtained from stochastic simulations of the process based on a Langevin equation approach, see Section~\ref{num simul}.

\subsection{Langevin equation approach. Numerical simulations}\label{num simul}

To verify the analytical solution obtained in the previous section, we perform numerical calculations, considering a system of Langevin equations~\cite{iomin jstat,lenzi njp}, and where resets, as a renewal process, can be easily performed, see Ref.~\cite{pal_pre2015}. The system of coupled Langevin equations reads
{\small{\begin{subequations}
\begin{align}
&x(t+\Delta t)=\left\{
            \begin{array}{ll}
              x(0),\, \textrm{with prob.} \: r\,\Delta t,
              \\ \\
              x(t) + \beta_1 A(y) B(z) \eta _x (t),\, \textrm{with prob.} \: (1-r\,\Delta t), 
            \end{array}
          \right.\label{lex}
\end{align}
\begin{align}
y(t+\Delta t)=\left\{
            \begin{array}{ll}
              y(0),\, \textrm{with prob.} \: r\,\Delta t,
              \\ \\
              y(t) + \beta_2 B(z) \eta _y (t),\, \textrm{with prob.} \: (1-r\,\Delta t), 
            \end{array}
          \right. \label{ley}
\end{align}
\begin{align}
z(t+\Delta t)=\left\{
            \begin{array}{ll}
              z(0),\, \textrm{with prob.} \: r\,\Delta t,
              \\ \\
              z(t) + \beta_3 \eta _z (t),\, \textrm{with prob.} \: (1-r\,\Delta t), \end{array}
          \right.
\end{align}
\end{subequations}
}}where $\beta_1$, $\beta_2$, $\beta_3$ are constants related to the diffusion coefficients $\mathcal{D}_1$, $\mathcal{D}_2$, $\mathcal{D}_3$, $\eta_x (t)$, $\eta_y (t)$, $\eta_z (t)$ are zero mean Gaussian noises ($\langle \eta _x (t) \rangle = 0,\: \langle \eta _y (t) \rangle = 0,\: \langle \eta _z (t) \rangle = 0$), $A(y)$ and $B(z)$ are functions introduced to mimic $\delta$-functions (see Refs.~\cite{iomin jstat,jstat2020}), and $r$ is the parameter of the Poisson process. To replicate the Dirac $\delta$-function, diffusion across the $x$ and $y$ directions is permitted in a narrow band of thickness $2\varepsilon$ along the $x$ and $y$ axes. As a result, the noise in Eqs.~(\ref{lex}) and (\ref{ley}) is multiplicative, however in Refs.~\cite{lenzi njp,jstat2020} the authors verified that the value $\varepsilon$ has no influence in the diffusive process, as long as $\varepsilon$ and the noise amplitudes $\beta_1$, $\beta_2$, $\beta_3$ are of the same order of magnitude. In our simulations, we have set $\varepsilon = \beta_x = \beta_y = \beta_z = 0.1$. The noises $\eta_x(t)$, $\eta_y(t)$, $\eta_z(t)$, were sampled from a Gaussian distribution $N(0,\Delta t)$. The time evolution of the diffusive particle is a renewal process, where each resetting event to $(x_0,y_0,z_0)$ renews the process at a Poisson rate $r$.

\begin{widetext}
 	
 	\begin{figure}
 		\includegraphics[width=0.95\textwidth]{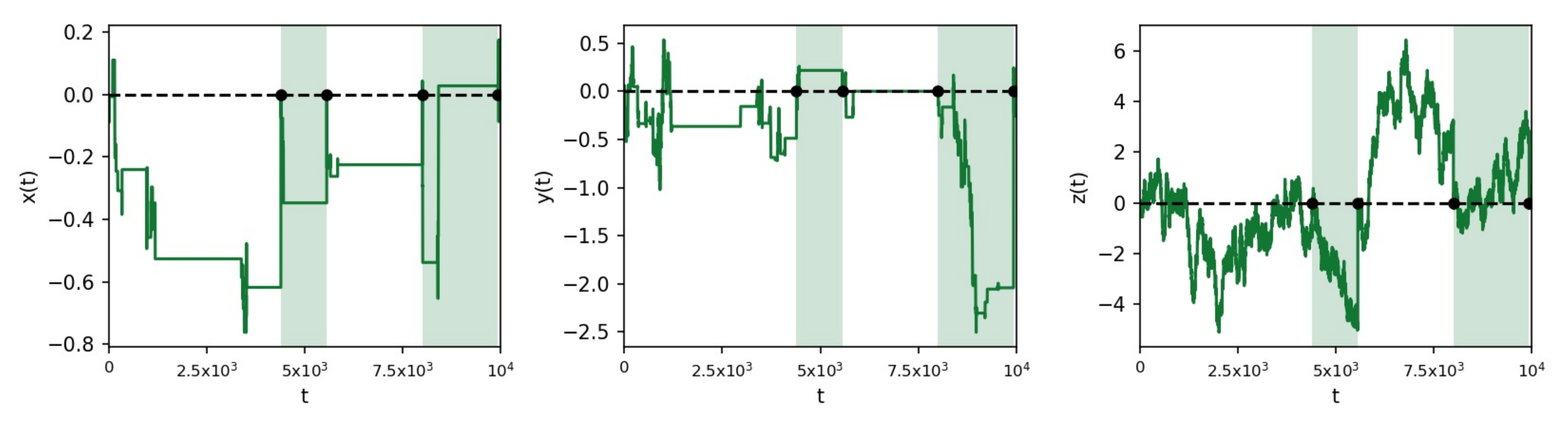}\caption{Trajectory along individual axes with global stochastic resetting to the initial position $(x_0,y_0,z_0)=(0,0,0)$ with rate $r=0.002$ obtained from a Langevin simulation of the process. The resetting events are represented by black dots. Dashed regions are introduced in order for these resetting events to be more visible. \label{AxisTrajectories_g}}
 	\end{figure}
 	
 \end{widetext}

 This effect of stochastic resetting is modeled by sampling a resetting time from an exponential distribution with parameter $r$ representing the time between two events in a Poisson point process. During this resetting time, the particle undergoes diffusion on the three-dimensional comb and resets at $(x_0,y_0,z_0)$ afterwards. Graphical representation of the simulations of particle trajectories along all directions is given in Fig.~\ref{AxisTrajectories_g}.
 
\begin{figure}
\centering{\includegraphics[width=0.5\textwidth]{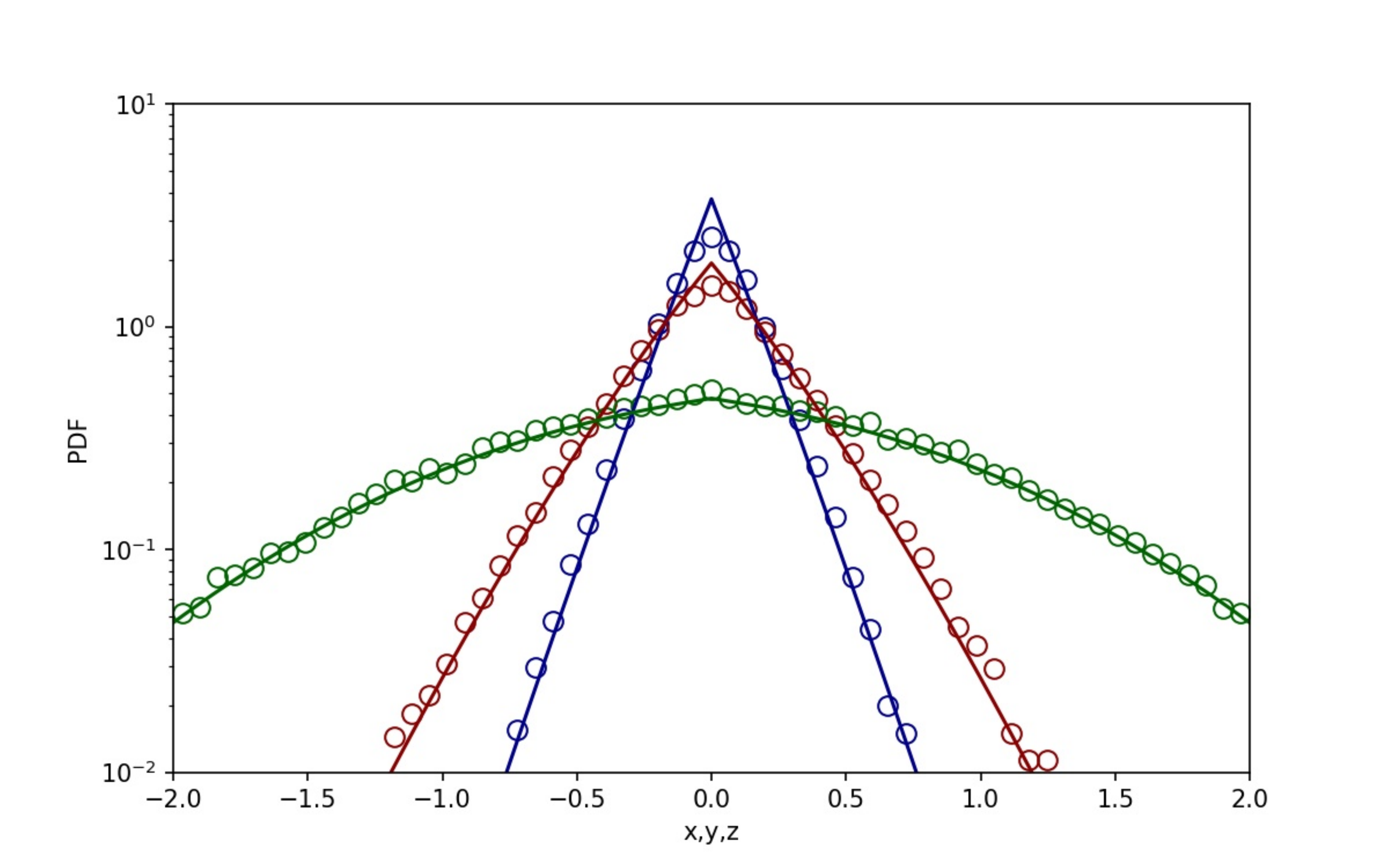}
\caption{Comparison of the analytical (solid lines) and simulation (dots) results for the marginal PDFs for global resetting with $r=0.002$ at $t=100$, and reset position $x_0=0$. More specifically, we show $p_{1}(x,t|0)$ (blue line), $p_{2}(y,t|0)$ (red line) and $p_{3}(z,t|0)$ (green line) for $\mathcal{D}_{1}=\mathcal{D}_{1}=\mathcal{D}_{3}=0.005$. }\label{figP1}}
\end{figure}

\begin{figure}
	\centering{\includegraphics[width=0.43\textwidth]{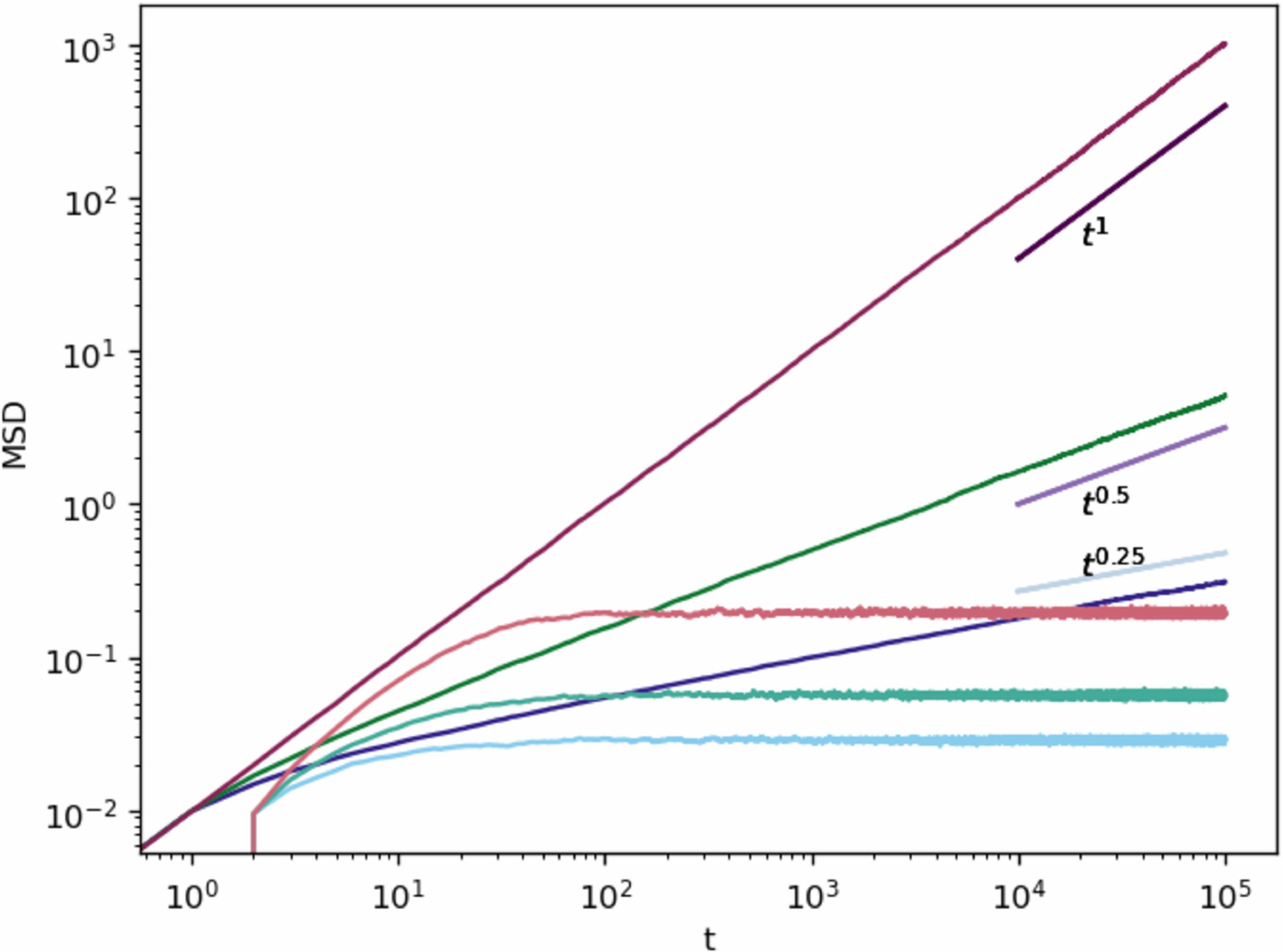}\caption{MSDs along all three axes for global resetting with rate $r=0.05$. Each color corresponds to an axis, $x$ (blue), $y$ (green), $z$ (red) with and without resetting.}\label{MSD-X}}
\end{figure}

Regarding the simulation of marginal PDFs and temporal evolution of the variance, ensembles of $5\times10^4$ particle positions were simulated considering a time step of $\Delta t=1$ across a time span of $10^5$ in order to observe convergence of the processes, with the MSD being calculated for each of the ensembles along all three directions
$\sigma^2 _x (t) = \langle (x(t) - \langle x(t) \rangle )^2 \rangle$, 
$\sigma^2 _y (t) = \langle (y(t) - \langle y(t) \rangle )^2 \rangle$, 
$\sigma^2 _z (t) = \langle (z(t) - \langle z(t) \rangle )^2 \rangle$. 
In Fig.~\ref{figP1} we give comparison of the analytical and simulation results for the marginal PDFs, where we use that $\beta_{i}=\sqrt{2\,\mathcal{D}_{i}}$, $i=\{1,2,3\}$, with $\Delta t=1$, see \cite{iomin jstat,pal_pre2015}. In Fig.~\ref{MSD-X} we show the simulated time evolution of the MSDs in the three directions. From the simulation results one can verify that they are in a very good agreement with the saturation values obtained analytically, if one uses that $2\,\mathcal{D}_{i}=\beta_{i}^{2}$. For more results on the corresponding PDFs obtained by the numerical simulations we refer to the {\it Supplemental material 1}.

\section{Resetting to the backbone}

Global resetting takes the particle to a particular position of the comb (with the coordinates
$(x_0, 0, 0)$ in Eq.~\eqref{diffusion eq 3Dcomb resetting}). However, it is only one of many possible mechanisms of resetting. Here, we proceed with a slightly softer resetting procedure, which takes the particle to the backbone. This resetting is applied to the $y$ and $z$ directions only, taking a walker being at $(x,y,z)$ to the point $(x,0,0)$. In this case, the governing equation reads
{\small{\begin{align}
&\frac{\partial }{\partial t}P(x,y,z,t|x_0,0,0) = L_{\mathrm{FP}}P(x,y,z,t|x_0,0,0) 
\nonumber\\&+r\delta(y)\delta(z)\int_{-\infty}^\infty dy'\int_{-\infty}^{\infty}dz'P(x,y',z',t|x_0,0,0),\nonumber\\
\label{EqFokkerPlanckResBackbone1}
\end{align}
}}which differs from Eq.~\eqref{diffusion eq 3Dcomb resetting} in the last term only. This difference results from the difference between the global resetting and resetting to the backbone. In the former case the particle is taken at the particular position $(x_0,0,0)$ as stated by the $\delta(x-x_0)\delta(y)\delta(z)$ term in Eq.~\eqref{diffusion eq 3Dcomb resetting}. However, in the latter case, considered here, the particle appears at $y=0,\ z=0$ but the $x$ position is not modified. Mathematically it can be written as the marginal distribution, the double integral term in Eq.~\eqref{EqFokkerPlanckResBackbone1}. From the Fourier-Laplace transformations, we arrive at the following PDF in the Fourier-Laplace space, see {\it Supplemental material 2} for details,
\begin{align}
&\hat{P}(k_x,k_y,k_z,s|x_0,0,0)\nonumber\\&
=\frac{1}{s}\times\frac{s\,(s+r)^{-3/4}}{s\,(s+r)^{-3/4}+\mathcal{D}_{1}\,k_x^2}\,e^{ik_x x_0}\nonumber\\
&\times \frac{(s+r)^{1/2}}{(s+r)^{1/2}+\mathcal{D}_{2}\,k_y^2}\times\frac{s+r}{s+r+\mathcal{D}_3\,k_z^2}.
\label{EqPropBackboneFL2}
\end{align}
From this equation, the marginal PDFs for the three axis can be straightforwardly obtained as done in the previous section:
\begin{equation}\label{fungers pdf1}
\hat{p}_1(k_x,s)=\frac{(s+r)^{-3/4}}{s\,(s+r)^{-3/4}+\mathcal{D}_1\,k_x^2}\,e^{ik_x x_0}, 
\end{equation}
\begin{equation}\label{fungers pdf2}
\hat{p}_2(k_y,s)=\frac{s^{-1}\,(s+r)^{1/2}}{(s+r)^{1/2}+\mathcal{D}_{2}\,k_y^2}, 
\end{equation}
\begin{equation}\label{fungers pdf3}
\hat{p}_3(k_z,s)=\frac{s^{-1}\,(s+r)}{(s+r)+\mathcal{D}_3\,k_z^2}. 
\end{equation}
Here we note that the corresponding equations, Eqs.~(\ref{fungers pdf2}) and (\ref{fungers pdf3}), for the marginal PDFs along the $y$ and $z$ directions are the same as in the case of global resetting, Eqs.~(\ref{P(kx,ky,kz,s) final2}) and (\ref{global p3}), respectively. Along the backbone the PDF is
\begin{equation}
\hat{p}_1(x,s)=\frac{1}{2\sqrt{\mathcal{D}_{1}}}s^{-1/2}(s+r)^{-3/8}\,e^{-\frac{s^{1/2}(s+r)^{-3/8}}{\sqrt{\mathcal{D}_{1}}}|x-x_0|},
\end{equation}
which is governed by the equation \begin{align}
{_{TC}}D_{r}^{1/4}p_{1}(x,t|x_0)=\mathcal{D}_{1}\frac{\partial^2}{\partial x^2}p_{1}(x,t|x_0),
\end{align}
where ${_{TC}}D_{b}^{\alpha}f(t)$ is the tempered fractional derivative (\ref{TC derivative}) of order $1/4$. 

The corresponding MSDs along $y$ and $z$ axes are the same as those for the case of global resetting, since the effect of the resetting in these two dimensions is equivalent for both scenarios. However, the dynamics on the $x$ axis change substantially as reflected in the MSD
\begin{equation}
\langle x^2(t)\rangle=x_0^2+2\,\mathcal{D}_{1}\,t^{1/4}E_{1,5/4}^{-3/4}(-r\,t).
\label{EqMSDxaxis2}
\end{equation}
Its asymptotics reads
\begin{align}
    \left\langle x^{2}(t)\right\rangle\sim x_{0}^{2}+2\,\mathcal{D}_{1}\left\lbrace\begin{array}{l l}
     & \frac{t^{1/4}}{\Gamma(5/4)}, \quad rt\ll1, \\ \\
     & r^{3/4}\,t, \quad rt\gg1.
\end{array}\right.
\end{align}.
The resetting mechanism studied in this section enhances the transport since it returns particles to the $x$ axis. Consequently, instead of the saturation of a stationary value for the MSD, one can see from Eq.~\eqref{EqMSDxaxis2} that in the long time limit, $\langle x^2(t)\rangle\sim t$, i.e. it scales diffusively. The short time limit scales as $\langle x^2(t)\rangle\sim t^{1/4}$, as in the case of global resetting. This means that we observe accelerating transport along the backbone, ranging from subdiffsuion to normal diffusion.

\begin{widetext}

\begin{figure}
\includegraphics[width=0.95\textwidth]{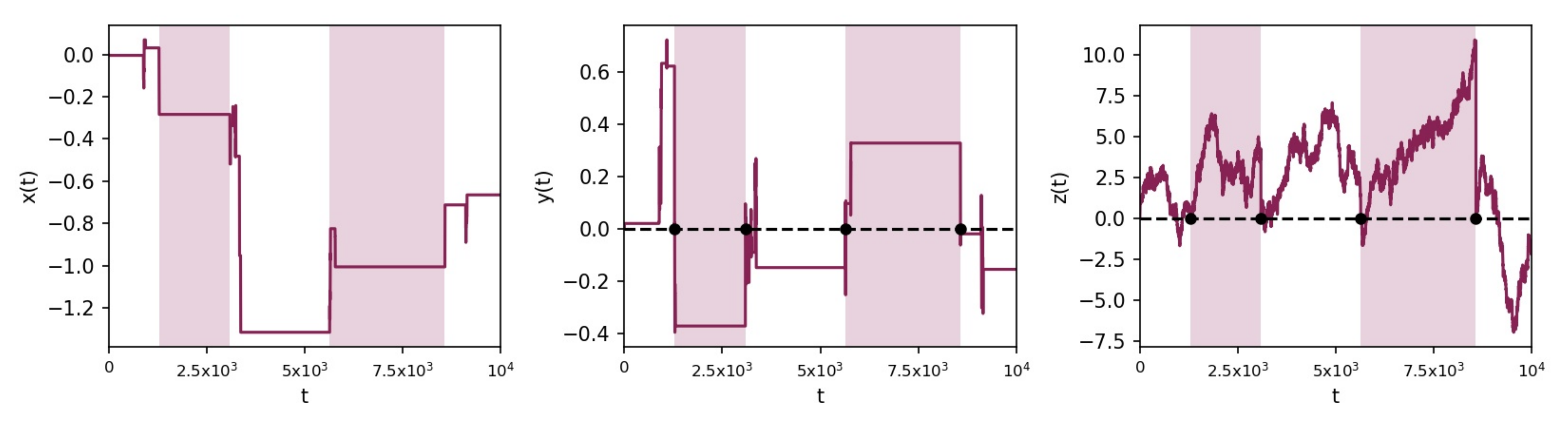}
\caption{Trajectory along individual axes with stochastic resetting to the backbone with rate $r=0.002$ obtained from a Langevin simulation of the process. The resetting events are represented by black dots. Note that there is no resetting along $x$ axis but the dashed regions are given only for indication when the resetting events along $y$ and $z$ axis occur.\label{AxisTrajectories_b}}
\end{figure}

\end{widetext}

\begin{figure}
\centering{\includegraphics[width=0.5\textwidth]{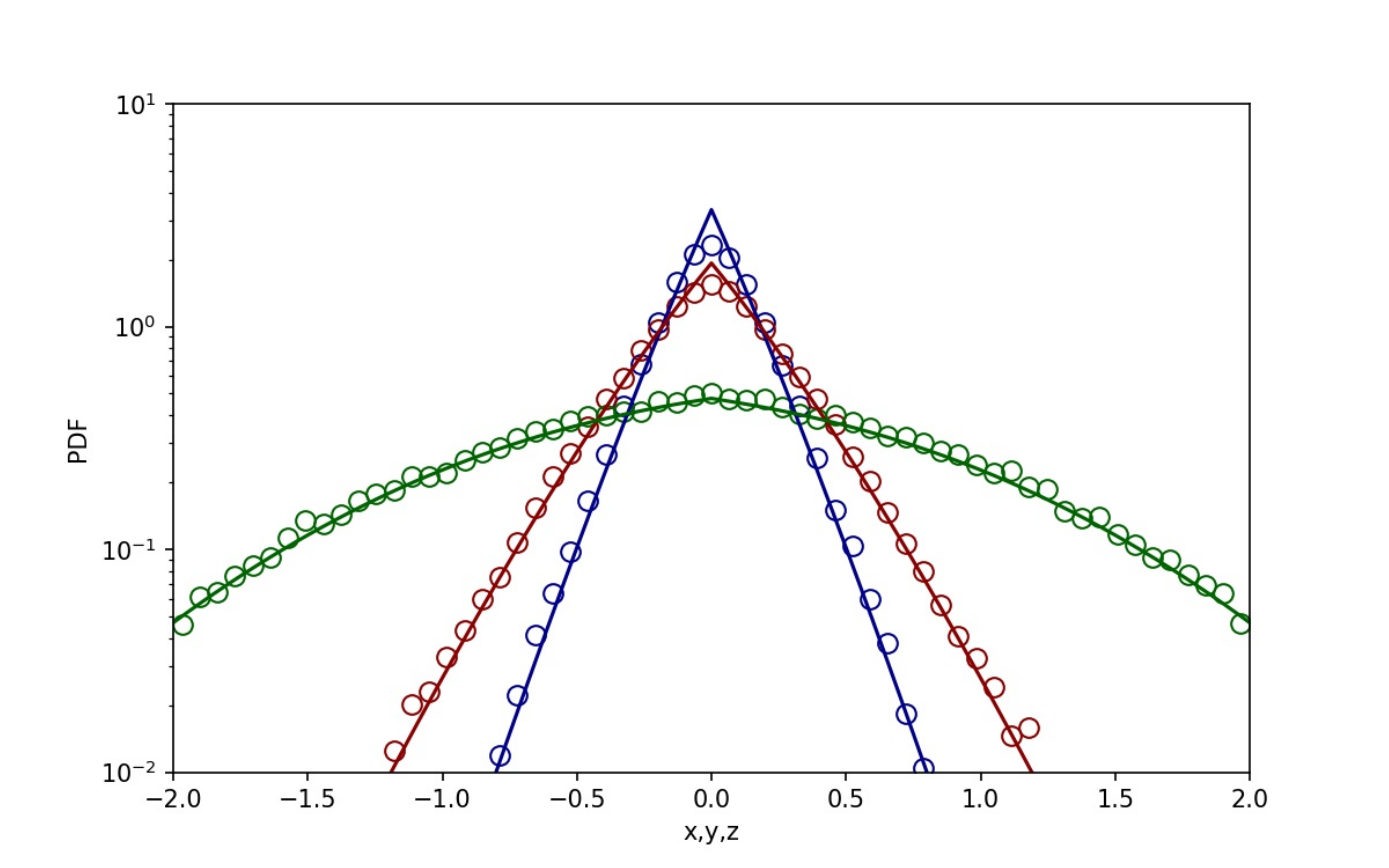}
\caption{Comparison of the analytical (solid lines) and simulation (dots) results for the marginal PDFs for resetting to the backbone with $r=0.002$ at $t=100$, and reset position $x_0=0$. We show $p_{1}(x,t|0)$ (blue line), $p_{2}(y,t|0)$ (red line) and $p_{3}(z,t|0)$ (green line) for $\mathcal{D}_{1}=\mathcal{D}_{2}=\mathcal{D}_{3}=0.005$.}\label{figP11}}
\end{figure}

\begin{figure}
\centering{\includegraphics[width=0.43\textwidth]{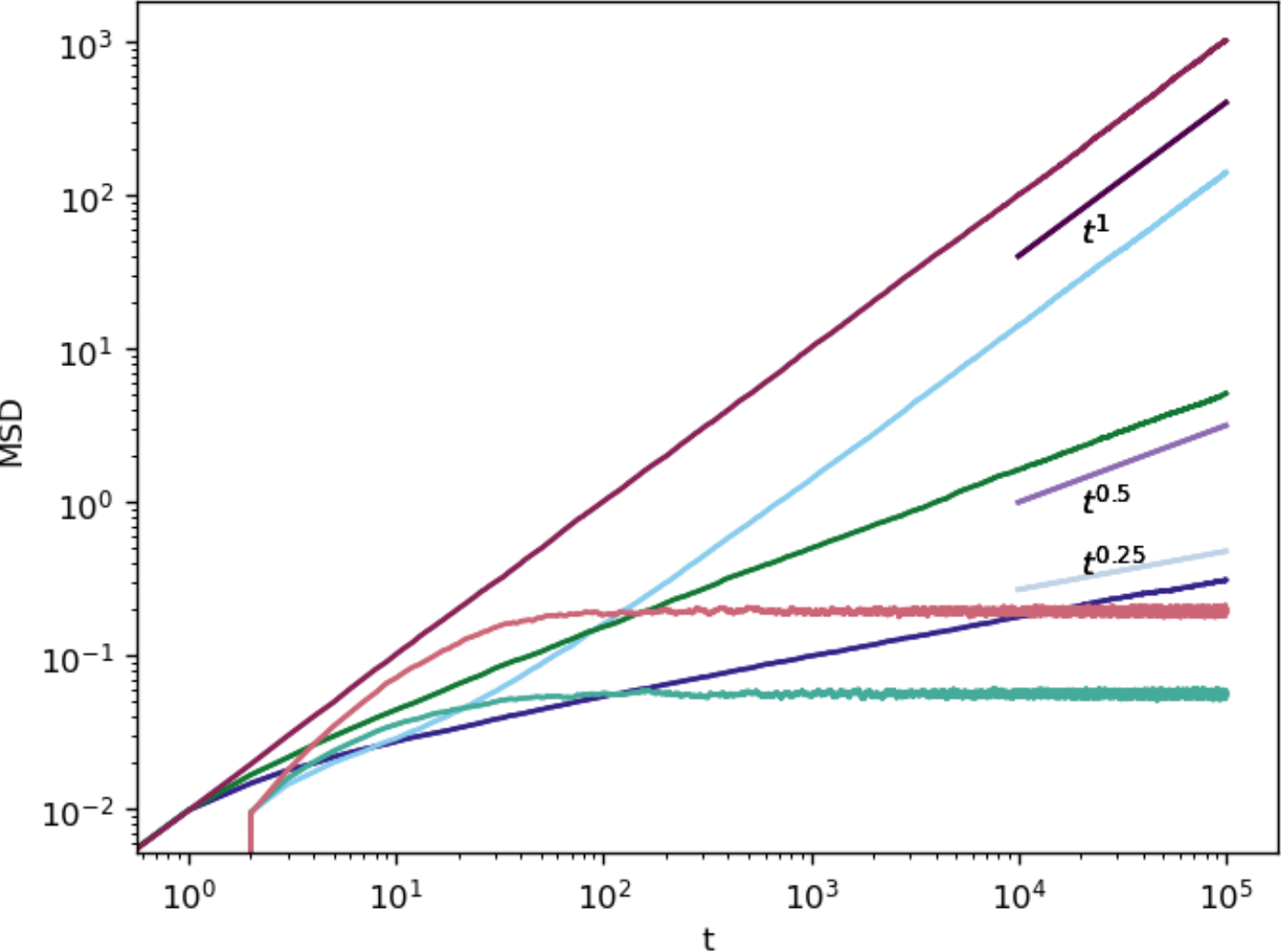}\caption{MSDs along all three axes for resetting to the backbone with rate $r=0.05$. The same color pattern as in Fig.~\ref{MSD-X} is used.}\label{MSD-X-b}}
\end{figure}

The numerical simulations of particle trajectories along all three directions, by using the Langevin equations approach, are shown in Fig.~\ref{AxisTrajectories_b}. We see that while in the $y$ and $z$ axis we observe recurrent returns to the origin, in the $x$ axis the motion does not return. Instead, it freely moves away from the origin. In Fig.~\ref{figP11} we give comparison of the analytical and simulation results for the marginal PDFs from where one observes excellent agreement between both approaches. Same parameters for $\varepsilon$, $\beta_{i}$ and $\mathcal{D}_{i}$ as in the case of global resetting are used. The analytical results has also been confirmed by simulation results of the MSDs given in Fig.~\ref{MSD-X-b}, where the $y$ and $z$ components of the MSD reach a stationary value while the MSD in the $x$ direction increases linearly. This is in agreement with the analytical results found above. More simulation results for the PDFs are given in the {\it Supplemental material 2}.


\section{Resetting to the main fingers}


Finally, we study the dynamics of the system when the resetting applies to particle located at any secondary finger along the $z$ axis and moves to main finger (axis $y$ in Fig.~\ref{Fig1}). In this case, the governing equation reads
{\small{\begin{align}
&\frac{\partial }{\partial t}P(x,y,z,t|x_0,0,0)=L_{\mathrm{FP}}P(x,y,z,t|x_0,0,0) 
\nonumber\\&-rP(x,y,z,t|x_0,0,0)+r\delta(z)\int_{-\infty}^{\infty}dz'P(x,y,z',t|x_0,0,0),
\label{EqFokkerPlanckResFingers1}
\end{align}
}}where the last term is now the marginal distribution in the variables $x$ and $y$. In the Fourier-Laplace space, the solution of the equation reads, see {\it Supplemental material 3} for details,
\begin{align}
&\hat{P}(k_x,k_y,k_z,s|x_0,0,0)\nonumber\\ &=\frac{1}{s}\times\frac{s^{1/2}\,(s+r)^{-1/4}}{s^{1/2}\,(s+r)^{-1/4}+\mathcal{D}_1\,k_x^2}\,e^{ik_x x_0}\nonumber\\&
\times\frac{s\,(s+r)^{-1/2}}{s\,(s+r)^{-1/2}+\mathcal{D}_2\,k_y^2}
\times \frac{s+r}{s+r+\mathcal{D}_3\,k_z^2}.
\label{EqPropFingersFL2}
\end{align}
This also yields the images of the marginal PDFs for the different axis. Eventually, we have
\begin{equation}
\hat{p}_1(k_x,s)=\frac{s^{-1/2}\,(s+r)^{-1/4}}{s^{1/2}\,(s+r)^{-1/4}+\mathcal{D}_1\,k_x^2}\,e^{ik_x x_0},
\end{equation}
\begin{equation}
\hat{p}_2(k_y,s)=\frac{(s+r)^{-1/2}}{s\,(s+r)^{-1/2}+\mathcal{D}_2\,k_y^2},
\end{equation}
\begin{equation}
\hat{p}_3(k_z,s)=\frac{s^{-1}\,(s+r)}{(s+r)+\mathcal{D}_3\,k_z^2}.
\end{equation}

From these expressions one obtains the corresponding MSDs, which are
\begin{equation}
\langle x^2(t)\rangle=x_0^2+2\,\mathcal{D}_{1}\,t^{1/4}E_{1,5/4}^{-1/4}(-r\,t), 
\label{EqMSDxaxis2}
\end{equation}
\begin{align}
\langle y^2(t)\rangle&=2\,\mathcal{D}_{2}\,t^{1/2}E_{1,3/2}^{-1/2}(-r\,t)\nonumber\\&=2\,\mathcal{D}_{2}\left[e^{-rt}\frac{t^{1/2}}{\Gamma(1/2)}+\frac{2rt+1}{2\sqrt{r}}\mathrm{erf}(\sqrt{rt})\right], 
\end{align}
and the MSD along the $z$ axis is the same as in the previous cases. Their asymptotics read
\begin{align}
    \left\langle x^{2}(t)\right\rangle\sim x_{0}^{2}+2\,\mathcal{D}_{1}\left\lbrace\begin{array}{l l}
     & \frac{t^{1/4}}{\Gamma(5/4)}, \quad rt\ll1,\\ \\
     & \frac{r^{1/4}\,t^{1/2}}{\Gamma(3/2)}, \quad rt\gg1,
\end{array}\right.
\end{align}
\begin{align}
    \left\langle y^{2}(t)\right\rangle\sim 2\,\mathcal{D}_{2}\left\lbrace\begin{array}{l l}
     & \frac{t^{1/2}}{\Gamma(3/2)}, \quad rt\ll1,\\ \\
     & r^{1/2}\,t, \quad rt\gg1,
\end{array}\right.
\end{align}
In this case, the MSD in the $x$-axis
behaves subdiffusively with $\langle x^2(t)\rangle \sim t^{1/4}$ as in the case with no resetting, and then it turns to $\langle x^2(t)\rangle \sim t^{1/2}$, which means an accelerating subdiffusive transport. Along the $y$-axis, the MSD scales as $\langle y^2(t)\rangle\sim t^{1/2}$ in the short time limit, and then it turns to linear dependence in time $\langle y^2(t)\rangle\sim t$. Along the $z$-axis the MSD from the normal diffusive behavior reaches a stationary value in the long time limit.

\begin{widetext}

\begin{figure}
\includegraphics[width=0.95\textwidth]{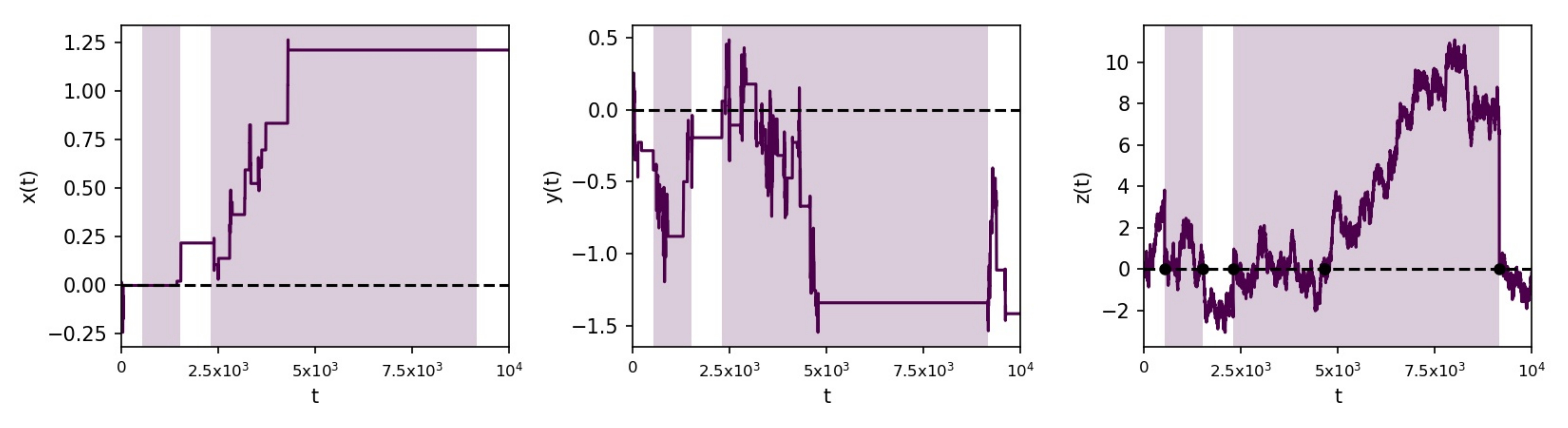}
\caption{Trajectory along individual axes with stochastic resetting to the fingers with rate $r=0.002$ obtained from a Langevin simulation of the process. The resetting events are represented by black dots. Note that there is no resetting along $x$ and $y$ axis but the dashed regions are given only for indication when the resetting events along $z$ axis occur.\label{AxisTrajectories_f}} 
\end{figure}

\end{widetext}

\begin{figure}
\centering{\includegraphics[width=0.5\textwidth]{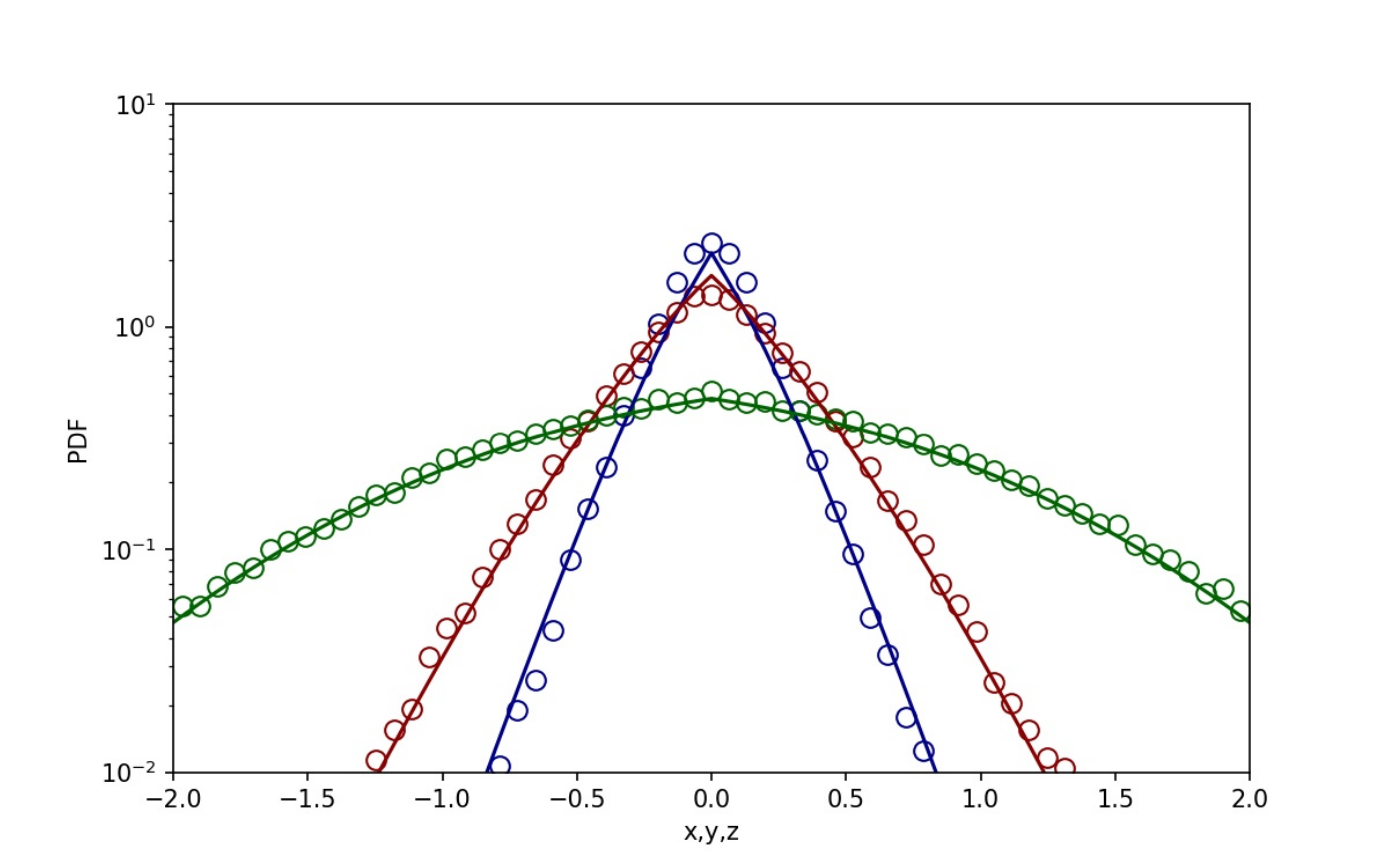}
\caption{Comparison of the analytical (solid lines) and simulation (dots) results for the marginal PDFs for resetting to the fingers with $r=0.002$ at $t=100$, and reset position $x_0=0$. We show $p_{1}(x,t|0)$ (blue line), $p_{2}(y,t|0)$ (red line) and $p_{3}(z,t|0)$ (green line) for $\mathcal{D}_{x}=\mathcal{D}_{y}=\mathcal{D}_{z}=1$.}\label{figP111}}
\end{figure}

\begin{figure}
\centering{\includegraphics[width=0.43\textwidth]{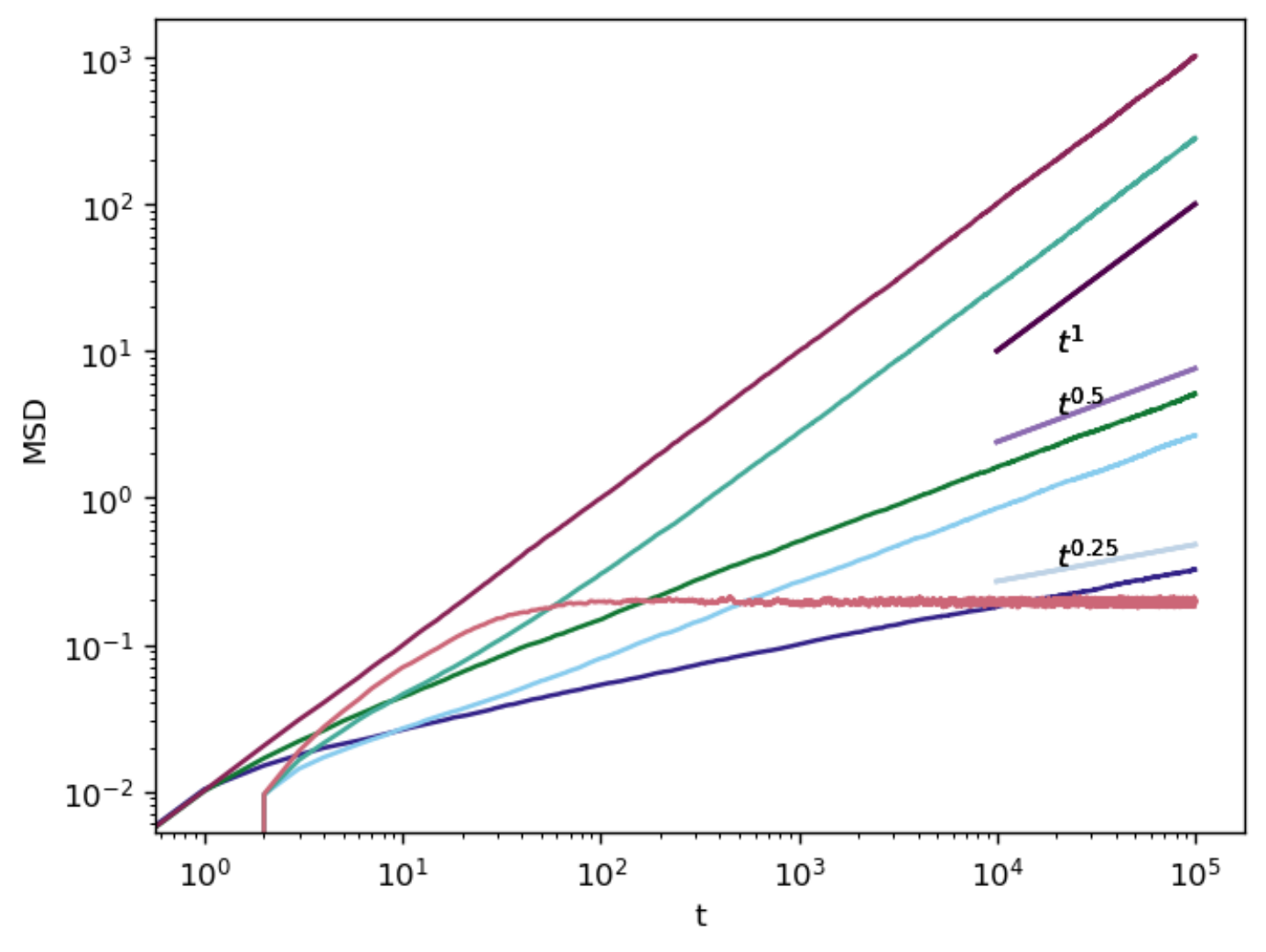}\caption{MSDs along all three axes for resetting to the fingers with rate $r=0.05$. The same color pattern as in Fig.~\ref{MSD-X} is used.}\label{MSD-X-F}}
\end{figure}

We also performed numerical simulations by using the Langevin equations approach. Same parameters for $\varepsilon$, $\beta_{i}$ and $\mathcal{D}_{i}$ as previous are used. The simulation results show very good agreement with the analytical results, see Figs.~\ref{AxisTrajectories_f}, \ref{figP111} and 
\ref{MSD-X-F}. For more details on analytical computation and simulation results see also {\it Supplemental material 3}.


\section{Remarks on two dimensional comb}

Here we note that the results obtained for the three-dimensional $xyz$-comb can be used for the two-dimensional $xy$-comb. The $y$ and $z$ axes in the three-dimensional comb would correspond to the $x$ and $y$ axes in the two-dimensional comb. Therefore, the results obtained for the $y$ and $z$ directions in the three-dimensional comb with resetting in the backbone correspond to the results for the $x$ and $y$ directions in the two-dimensional comb with global resetting. Furthermore, the results obtained for the $y$ and $z$ directions in the three-dimensional comb with resetting in the main fingers correspond to the results for the $x$ and $y$ directions in the two-dimensional comb with resetting in the fingers.

\section{Remark on the three dimensional comb geometry}

The topological (comb) constraint of the transport properties of both two- and three-dimensional combs should be discussed as well. To that end, let us understand the role of the $\delta(y)$ and $\delta(z)$ functions in the highly inhomogeneous diffusion coefficients in Eq.~\eqref{diffusion eq 3Dcomb}. One should recognize that the singularity of the $x$ and $y$ components of the diffusion tensor is the intrinsic transport property of the comb model~\eqref{diffusion eq 3Dcomb}. Note that this singularity of the diffusion coefficients relates to a non-zero flux along the $x$ backbone and $y$ fingers, and for the two dimensional case it was discussed in Refs.~\cite{dvork2009,LZ98,IoZaPf16,SaIoMe16}. Here, we extend the arguments of Refs.~\cite{IoZaPf16,SaIoMe16} for the three dimensional case of Eq.~\eqref{diffusion eq 3Dcomb}. Let us consider the Liouville equation
\begin{equation}\label{liouville eq}
\frac{\partial}{\partial t}P+\mathrm{div}\,\mathbf{j}=0,
\end{equation}
where the three dimensional current $\mathbf{j}=(j_x,\,j_y,\,j_z)$ describes Markov processes in Eq.~\eqref{diffusion eq 3Dcomb}.
In this case, the three-dimensional current reads
\begin{subequations}\label{JxJyJz}
\begin{align}
& j_x=-D_x(y,z)\,\frac{\partial}{\partial x}P(x,y,z,t), \label{JxJy-x}\\
& j_y=-D_y(z)\,\frac{\partial}{\partial y}P(x,y,z,t),
\label{JxJy-y} \\
& j_z=-D_z\,\frac{\partial}{\partial z}P(x,y,z,t). \label{JxJy-z}
\end{align}
\end{subequations}
Here, we take a general diffusivity function in the $x$ and $y$ directions $D_x(y,z)$ and $D_y(z)$, respectively (instead of $\mathcal{D}_x\,\delta(y)\delta(z)$ and
$\mathcal{D}_y\,\delta(z)$ in Eq.~\eqref{diffusion eq 3Dcomb}).
Therefore, Eq.~(\ref{liouville eq}) together with Eqs.~\eqref{JxJyJz}, can be regarded as the three-dimensional non-Markovian master equation.

Integrating Eq.~(\ref{liouville eq}) with respect to $y$ and $z$ from $-\epsilon/2$ to $\epsilon/2$: $\int_{-\epsilon/2}^{\epsilon/2}dy\dots$ and $\int_{-\epsilon/2}^{\epsilon/2}dz\dots$, one obtains for the l.h.s. of the equation, after application of the middle point theorem,
$\epsilon^2\,\frac{\partial}{\partial t}P(x,y=0,z=0,t)$,
which is exact in the limit $\epsilon\rightarrow 0$. This term can be neglected in this limit $\epsilon\rightarrow 0$. Considering integration of the r.h.s. of the equation, one should bear
in mind that this procedure is artificial and its implementation needs some care. First we consider the currents outside of the $\epsilon$ vicinity of the $x$ backbone. In this case, according to the comb geometry, $j_x=0$ and we consider a two-dimensional $y-z$ comb. Therefore, we perform integration with respect to $z$ only. From Eq.~\eqref{JxJy-z} we obtain that this term responsible for the transport in the $z$ direction reads
$$\mathcal{D}_z \Big[P'(x,y,z,t)\big|_{z=\tfrac{\epsilon}{2}}- \\ P'(x,y,z,t)\big|_{z=-\tfrac{\epsilon}{2}}\Big] \, ,$$
where prime means derivative with respect to $z$. This corresponds to the two outgoing fluxes from the $y$ fingers in the $\pm z$ directions: $F_{z}(+)+F_{z}(-)$. The transport in the $y$ direction, after integration, is $$\epsilon\,D_y(z\rightarrow 0)\,\partial_y^2 P(x,y,z=0,t)\equiv F_{y}\, .$$ It should be stressed that the second derivative over $y$,
presented in the form $$\epsilon\,\frac{\partial^2}{\partial y^2}P=
\left[\frac{\partial}{\partial y}P(y+\epsilon/2) -\frac{\partial}{\partial y}P(y-\epsilon/2)\right]\, ,$$ ensures both incoming and outgoing fluxes for $F_{y}$ along the $y$ direction at a point $y$. Following the Kirchhoff's law, we have $F_{y}+F_{z}(+)+F_{z}(-)=0$ for every point $y$ and at $z=0$. Function $F_{y}$ contains both incoming and outgoing fluxes of the probability, while $F_{z}(+)$ and $F_{z}(-)$ are both outgoing probability fluxes. If the latter outgoing fluxes are not zero, the flux $F_{y}$ has to be nonzero as well: $F_{y}\neq 0$, as containing an incoming flux. Therefore, $\epsilon\,D_y(z\rightarrow 0)\neq 0$. Taking $D_y(z)=\frac{1}{\pi}\frac{\epsilon\mathcal{D}_y}{y^2+\epsilon^2}$, one obtains in the limit $\epsilon\rightarrow 0$ a nonzero flux $F_{y}$ with $D_y(z)=\mathcal{D}_y\,\delta(z)$, which is the diffusion coefficient in the $y$ direction in Eq.~\eqref{liouville eq}.

Now we perform integration in the $\epsilon$ vicinity of the $x$ backbone, where we take into account the singularity of the $y$ component of diffusion coefficient, which is $\mathcal{D}_y\,\delta(z)$. We also admit that integration of the $j_z$ current in Eq.~\eqref{JxJy-z} with respect to $y$ yields zero. Therefore, integration with respect to $y$ and $z$ yields from Eq.~\eqref{JxJy-y} the term responsible for the transport in the $y$ direction as follows
$$\mathcal{D}_y\Big[P'(x,y,z=0,t)\big|_{y=\tfrac{\epsilon}{2}}-P'(x,y,z=0,t)\big|_{y=-\tfrac{\epsilon}{2}}\Big] \, .$$Here prime means derivative with respect to $y$. This corresponds to the two outgoing fluxes from the backbone in the $\pm y$ directions: $F_{y}(+)+F_{y}(-)$.
The transport along the $x$ direction, after 
integration of Eq.~\eqref{JxJy-x}, is
\begin{multline*}
\epsilon^2\,D(y\rightarrow 0,z\rightarrow 0)\,
\frac{\partial^{2}}{\partial x^{2}}P(x,y=0,z=0,t)= \\
F_{x}(x+\epsilon)+F_{x}(x-\epsilon)\, .
\end{multline*}
In complete analogy with the $y$ coordinate, 
the second derivative with respect to $x$,
presented in the form $$\epsilon\,\frac{\partial^{2}}{\partial x^{2}}P=\left[\frac{\partial}{\partial x}P(x+\epsilon/2)-\frac{\partial}{\partial x}P(x-\epsilon/2)\right]$$ as $\epsilon\rightarrow 0$, ensures both incoming and outgoing fluxes for $F_{x}$ along the $x$
direction at a point $x$. Again, after the integration, the Liouville equation is a kind of the Kirchhoff's law: $F_{x}(+)+F_{x}(-)+F_{y}(+)+F_{y}(-)=0$
for each point $x$ and at $y=0$. Note, that the flax in the $z$ direction is zero due to the integration with respect to $y$. Since outgoing fluxes are not zero, $j_x\neq 0$ and correspondingly the flux $F_{x}\equiv F_{x}(+)+F_{x}(-)$ has to be nonzero as well: $F_{x}(\pm)\neq 0$. Therefore, $\epsilon^2\,D(y\rightarrow 0,z\rightarrow 0)\neq 0$.
Now, taking the diffusion coefficient in the form $D(y,z)=\frac{1}{\pi}\frac{\epsilon\mathcal{D}}{y^2+\epsilon^2}\cdot
\frac{1}{\pi}\frac{\epsilon\mathcal{D}}{z^2+\epsilon^2} $, one obtains in the limit $\epsilon\rightarrow 0$ a nonzero flux $F_{x}$ with
$D(y,z)=\mathcal{D}_x\delta(y)\delta(z)$, which is the diffusion coefficient in the $x$ direction in Eqs.~\eqref{diffusion eq 3Dcomb}, (\ref{liouville eq}) and \eqref{JxJy-x}.

\section{Summary}

We have studied the dynamics of a particle, which diffusing  in a three-dimensional heterogeneous comb-like structure performs different types of resets. The hierarchical structure of the three-dimensional comb allows to study different resetting mechanisms that generate a wide variety of dynamics depending on the strength of the resetting mechanism. In particular, we studied thee types of resets in the three-dimensional comb and their influence on the dynamics of the MSD, and we found that at the short time there is no influence on the transport exponents, which remain the same as in the case without resetting. However, at the long time limit, the system is strongly affected by the resetting process which leads a change in the transport exponents.

We studied three kinds of resetting: global resetting of a particle from any point on the comb to a fixed point at $(x,y,z)=(x_0,0,0)$ and two kinds of softer resetting, where two coordinates $(y=0,z=0)$ and one coordinate $(z=0)$ are fixed. When resets are global, the MSDs in $x$, $y$ and $z$ directions reach constant values exhibiting stochastic localization, i.e. a non-equilibrium steady state is reached. This result is in complete agreement with the results observed for the dynamics of walkers with constant rate resetting recently studied in the literature \cite{maj1,axel}. For a softer version of resetting consisting with two fixed coordinates, the walker returns to any positions at the backbone. It means that if the position of the walker before the resetting is $(x,y,z)$, then after the reset it is $(x,0,0)$. In this case, the dynamics for the $y$ and $z$ axis remains the same as in the global resetting case, since the effect of the resetting to these two coordinates is the same. However, in the $x$ direction, the resetting enhances the motion: it becomes subdiffusive $\left\langle x^2(t)\right\rangle\sim t^{1/4}$ for short times and then normal diffusion takes place for the long time scale. The latter regime results from the fact that the mean waiting time to stay in $(y,z)$ fingers becomes finite due to resetting. Indeed, the reset time PDF now plays the role of a waiting time PDF for the motion along the backbone ($x$ direction). Since the reset time PDF is exponential (i.e., constant rate resetting or Markovian resetting process), the motion in the $x$ direction becomes diffusive in the long time limit. For the softer resetting with one fixed $z$-coordinate, a stationary regime takes place in the $z$ fingers only. In the $x$ and $y$ directions the transport is enhanced with the MSDs behaving as follows: $\langle x^2(t)\rangle \sim t^{1/2}$ and $\langle y^2(t)\rangle \sim t$. The obtained diffusion equations for the marginal PDFs shed light on the physical relevance of usage of the regularized Prabhakar derivative in diffusion theory. They could describe diffusion processes on comb structures with stochastic resetting.

\section*{Acknowledgment}
{This research was partially supported by Grant No. CGL2016-78156-
C2-2-R (VM and AM). TS was supported by the Alexander von Humboldt Foundation.}

\bigskip

\newpage

\appendix

\begin{widetext}

\section*{Supplemental material: Stochastic resetting on comb-like structures, \ V. Domazetoski, A. Mas\'o-Puigdellosas, T. Sandev, V. M\'endez, A. Iomin and L. Kocarev}

\subsection{Solutions for diffusion with global resetting}\label{app1}

The diffusion equation in a three-dimensional comb with global resetting can be represented by
\begin{align}\label{diffusion eq 3Dcomb resettingApp}
\frac{\partial}{\partial t}P(x,y,z,t|x_0,0,0)&=\mathcal{D}_{x}\delta(y)\delta(z)\frac{\partial^2}{\partial x^2}P(x,y,z,t|x_0,0,0)+\mathcal{D}_{y}\delta(z)\frac{\partial^2}{\partial y^2}P(x,y,z,t|x_0,0,0)\nonumber\\&+\mathcal{D}_{z}\frac{\partial^2}{\partial z^2}P(x,y,z,t|x_0,0,0)-rP(x,y,z,t|x_0,0,0)+r\delta(x-x_0)\delta(y)\delta(z).
\end{align}
Here we consider the initial condition $P(x,y,z,t=0|x_0)=\delta(x-x_0)\delta(y)\delta(z)$. The Laplace transformation of Eq.~(\ref{diffusion eq 3Dcomb resettingApp}) reads
\begin{align}\label{diffusion eq 3Dcomb resetting laplace}
s\hat{P}(x,y,z,s|x_0,0,0)-\delta(x-x_0)\delta(y)\delta(z)&=\mathcal{D}_{x}\delta(y)\delta(z)\frac{\partial^2}{\partial x^2}\hat{P}(x,y,z,s|x_0,0,0)+\mathcal{D}_{y}\delta(z)\frac{\partial^2}{\partial y^2}\hat{P}(x,y,z,s|x_0,0,0)\nonumber\\&+\mathcal{D}_{z}\frac{\partial^2}{\partial z^2}\hat{P}(x,y,z,s|x_0,0,0)-r\hat{P}(x,y,z,s|x_0,0,0)+\frac{r}{s}\delta(x-x_0)\delta(y)\delta(z).
\end{align}
Then, by Fourier transform with respect to $x$, $y$ and $z$ we obtain
\begin{align}\label{diffusion eq 3Dcomb resetting laplace fourier3}
s\hat{P}(k_x,k_y,k_z,s|x_0,0,0)-e^{\imath k_x x_0}&=-\mathcal{D}_{x}k_{x}^{2}\hat{P}(k_x,y=0,z=0,s|x_0,0,0)-\mathcal{D}_{y}k_{y}^{2}\hat{P}(k_x,k_y,z=0,s|x_0,0,0)\nonumber\\&-\mathcal{D}_{z}k_{z}^{2}\hat{P}(k_x,k_y,k_z,s|x_0,0,0)-r\hat{P}(k_x,k_y,k_z,s|x_0,0,0)+\frac{r}{s}e^{\imath k_x x_0}.
\end{align}
from where it follows
\begin{align}\label{diffusion eq 3Dcomb resetting laplace fourier3}
\hat{P}(k_x,k_y,k_z,s|x_0,0,0)=\frac{s^{-1}(s+r)e^{\imath k_x x_0}}{(s+r+\mathcal{D}_{z}k_{z}^{2})}-\frac{\mathcal{D}_{x}k_{x}^{2}\hat{P}(k_x,y=0,z=0,s|x_0,0,0)}{(s+r+\mathcal{D}_{z}k_{z}^{2})}-\frac{\mathcal{D}_{y}k_{y}^{2}\hat{P}(k_x,k_y,z=0,s|x_0,0,0)}{(s+r+\mathcal{D}_{z}k_{z}^{2})},
\end{align}
The inverse Fourier transform with respect to $k_z$ yields
\begin{align}\label{diffusion eq 3Dcomb resetting laplace fourier3invZ}
\hat{P}(k_x,k_y,z,s|x_0,0,0)&=\frac{(s+r)^{-1/2}}{2\sqrt{\mathcal{D}_{z}}}\left[s^{-1}(s+r)e^{\imath k_x x_0}-\mathcal{D}_{x}k_{x}^{2}\hat{P}(k_x,y=0,z=0,s|x_0,0,0)\right.\nonumber\\&\left.-\mathcal{D}_{y}k_{y}^{2}\hat{P}(k_x,k_y,z=0,s|x_0,0,0)\right]e^{-\sqrt{\frac{s+r}{\mathcal{D}_{z}}}|z|},
\end{align}
from where we find $\hat{P}(k_x,k_y,z=0,s|x_0,0,0)$,
\begin{align}\label{diffusion eq 3Dcomb resetting laplace fourier3invZ=0}
\hat{P}(k_x,k_y,z=0,s|x_0,0,0)&=\frac{1}{\mathcal{D}_{y}}\frac{s^{-1}(s+r)e^{\imath k_x x_0}-\mathcal{D}_{x}k_{x}^{2}\hat{P}(k_x,y=0,z=0,s|x_0,0,0)}{\frac{2\sqrt{\mathcal{D}_{z}}}{\mathcal{D}_{y}}(s+r)^{1/2}+k_{y}^{2}}.
\end{align}
By applying the inverse Fourier transform with respect to $k_y$, we have
\begin{align}\label{diffusion eq 3Dcomb resetting laplace fourier3invZ=0invY}
\hat{P}(k_x,y,z=0,s|x_0,0,0)&=\frac{(s+r)^{-1/4}}{2\sqrt{2\mathcal{D}_{y}\sqrt{D}_{z}}}\left[s^{-1}(s+r)e^{\imath k_x x_0}-\mathcal{D}_{x}k_{x}^{2}\hat{P}(k_x,y=0,z=0,s|x_0,0,0)\right]e^{-\sqrt{\frac{2\sqrt{\mathcal{D}_{z}}}{\mathcal{D}_{y}}}},
\end{align}
and thus
\begin{align}\label{diffusion eq 3Dcomb resetting laplace fourier3invZ=0invY=0}
\hat{P}(k_x,y=0,z=0,s|x_0,0,0)&=\frac{1}{\mathcal{D}_{x}}\frac{s^{-1}(s+r)e^{\imath k_x x_0}}{\frac{2\sqrt{2\mathcal{D}_{y}\sqrt{\mathcal{D}_{z}}}}{\mathcal{D}_{x}}(s+r)^{1/4}+k_{x}^{2}}.
\end{align}

Therefore, we finally obtain
\begin{align}\label{P(kx,ky,kz,s) final0}
\hat{P}(k_x,k_y,k_z,s|x_0,0,0)=s^{-1}\frac{(s+r)^{1/4}}{(s+r)^{1/4}+\frac{\mathcal{D}_{x}}{2\sqrt{2\mathcal{D}_{y}\sqrt{\mathcal{D}_{z}}}}k_{x}^{2}}\times\frac{(s+r)^{1/2}}{(s+r)^{1/2}+\frac{\mathcal{D}_{y}}{2\sqrt{\mathcal{D}_{z}}}k_{y}^{2}}\times\frac{(s+r)}{(s+r)+\mathcal{D}_{z}k_{z}^{2}}\times e^{\imath k_{x}x_{0}}.
\end{align}

After we find the marginal PDFs we will analyze the mean squared displacements along all three directions, $\left\langle x^{2}(t)\right\rangle=\int_{-\infty}^{\infty}x^{2}\,p_{1}(x,t|x_0)\,dx$, $\left\langle y^{2}(t)\right\rangle=\int_{-\infty}^{\infty}y^{2}\,p_{2}(y,t|0)\,dy$, and $\left\langle y^{2}(t)\right\rangle=\int_{-\infty}^{\infty}z^{2}\,p_{3}(z,t|0)\,dz$. Therefore, we have
\begin{align}
    \left\langle x^{2}(t)\right\rangle&=2\left(\frac{\mathcal{D}_{x}}{2\sqrt{2\mathcal{D}_{y}\sqrt{\mathcal{D}_{z}}}}\right)\mathcal{L}^{-1}\left[\frac{s^{-1}}{(s+r)^{1/4}}\right]=2\left(\frac{\mathcal{D}_{x}}{2\sqrt{2\mathcal{D}_{y}\sqrt{\mathcal{D}_{z}}}}\right)t^{1/4}E_{1,5/4}^{1/4}(-rt)\nonumber\\&=2\left(\frac{\mathcal{D}_{x}}{2\sqrt{2\mathcal{D}_{y}\sqrt{\mathcal{D}_{z}}}}\right)t^{1/4}\left[\frac{1}{\sqrt[4]{rt}}-\frac{\textrm{E}_{3/4}(rt)}{\Gamma(1/4)}\right],
\end{align}
where
\begin{equation}\label{ML three a}
E_{\alpha,\beta}^{\delta}(z)=\sum_{k=0}^{\infty}\frac{(\delta)_k}{\Gamma(\alpha k+\beta)}\frac{z^k}{k!}
\end{equation}
is the three parameter Mittag-Leffler function, and $(\delta)_k=\Gamma(\delta+k)/\Gamma(\delta)$ is the Pochhammer symbol,
$\textrm{E}_{n}(z)=\int_{1}^{\infty}\frac{e^{-zt}}{t^{n}}\,dt$ is the exponential integral function,
\begin{align}
    \left\langle y^{2}(t)\right\rangle=2\left(\frac{\mathcal{D}_{y}}{2\sqrt{\mathcal{D}_{z}}}\right)\mathcal{L}^{-1}\left[\frac{s^{-1}}{(s+r)^{1/2}}\right]=2\left(\frac{\mathcal{D}_{y}}{2\sqrt{\mathcal{D}_{z}}}\right)t^{1/2}E_{1,3/2}^{1/2}(-rt)=2\left(\frac{\mathcal{D}_{y}}{2\sqrt{\mathcal{D}_{z}}}\right)\frac{\mathrm{erf}(\sqrt{rt})}{\sqrt{r}},
\end{align}
where $\mathrm{erf}(z)=\frac{2}{\sqrt{\pi}}\int_{0}^{z}e^{-t^2}\,dt$ gives the error function, and
\begin{align}
    \left\langle z^{2}(t)\right\rangle=2\,\mathcal{D}_{z}\,\mathcal{L}^{-1}\left[\frac{s^{-1}}{s+r}\right]=\frac{2\,\mathcal{D}_{z}}{r}\left[1-e^{-rt}\right].
\end{align}
The Laplace transform of the three parameter Mittag-Leffler function (\ref{ML three a}) reads
\begin{equation}\label{ML three Laplace}
\mathcal{L}\left[t^{\beta-1}E_{\alpha,\beta}^{\delta}(\pm at^{\alpha})\right](s)=\frac{s^{\alpha\delta-\beta}}{\left(s^{\alpha}\mp a\right)^{\delta}}, \quad \Re(s)>|a|^{1/\alpha},
\end{equation}
and its asymptotic behaviors are given by
\begin{equation}\label{GML asympt}
E_{\alpha,\beta}^{\gamma}(-z^{\alpha})\simeq\left\lbrace\begin{array}{ll}
     \frac{1}{\Gamma(\beta)}\exp\left(-\gamma\frac{\Gamma(\beta)}{\Gamma(\alpha+\beta)}z^{\alpha}\right), \quad z\ll1.  \\
     \frac{z^{-\alpha\gamma}}{\Gamma(\beta-\alpha\gamma)}, \quad z\gg1.
\end{array}\right.
\end{equation}

In Figs.~\ref{PDF-globalX0}, 
and \ref{PDF-globalX}, 
we give graphical representations of the PDFs of the process without resetting and under global resetting. The results are obtained by numerical simulations similar to the ones elaborated in Sec.~\ref{num simul}, with different values of the ensemble size of $2 \times 10^4$ and the time span of $5 \time 10^4$.

\begin{figure}
\centering{(a) \includegraphics[width=0.45\textwidth]{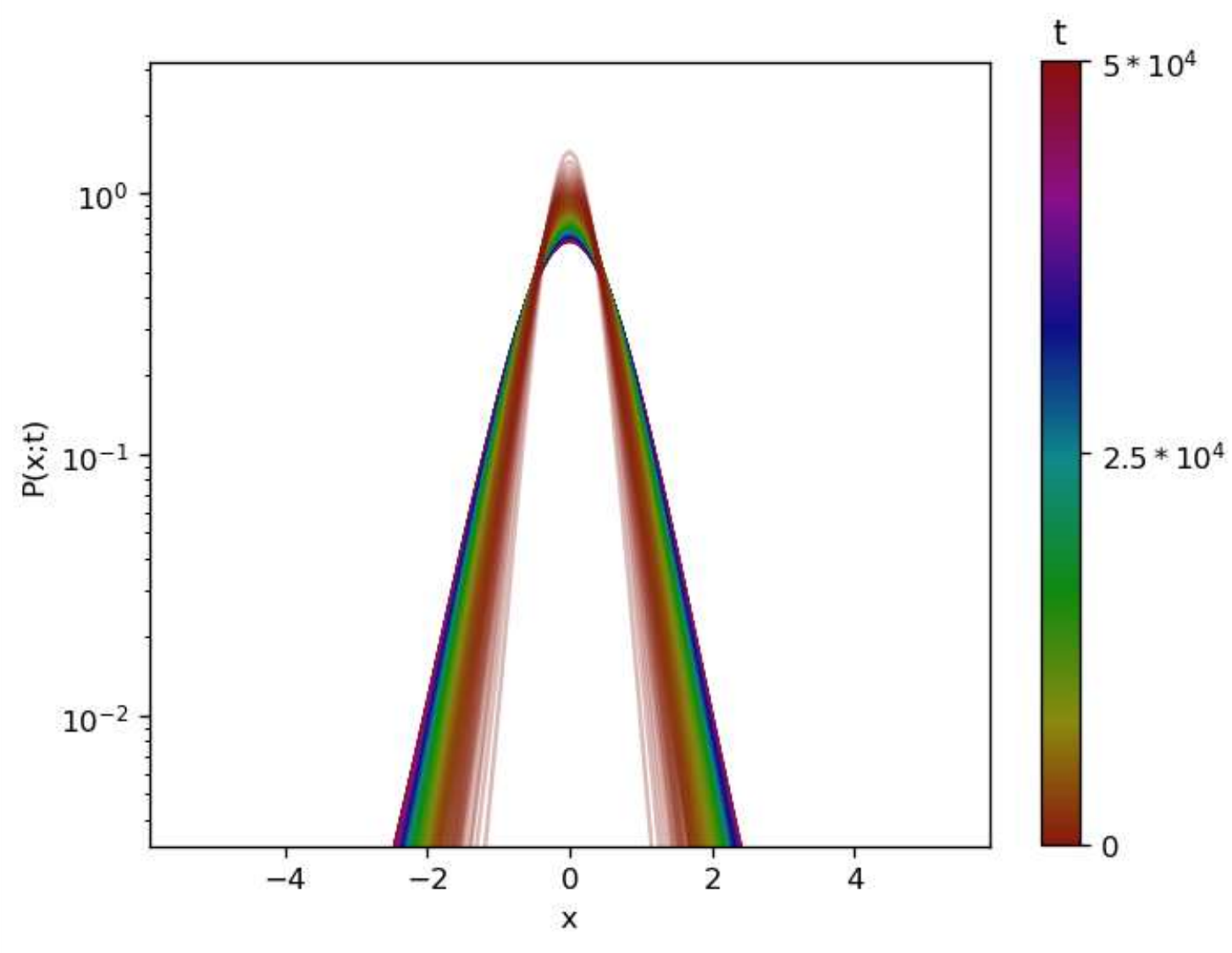} (b) \includegraphics[width=0.45\textwidth]{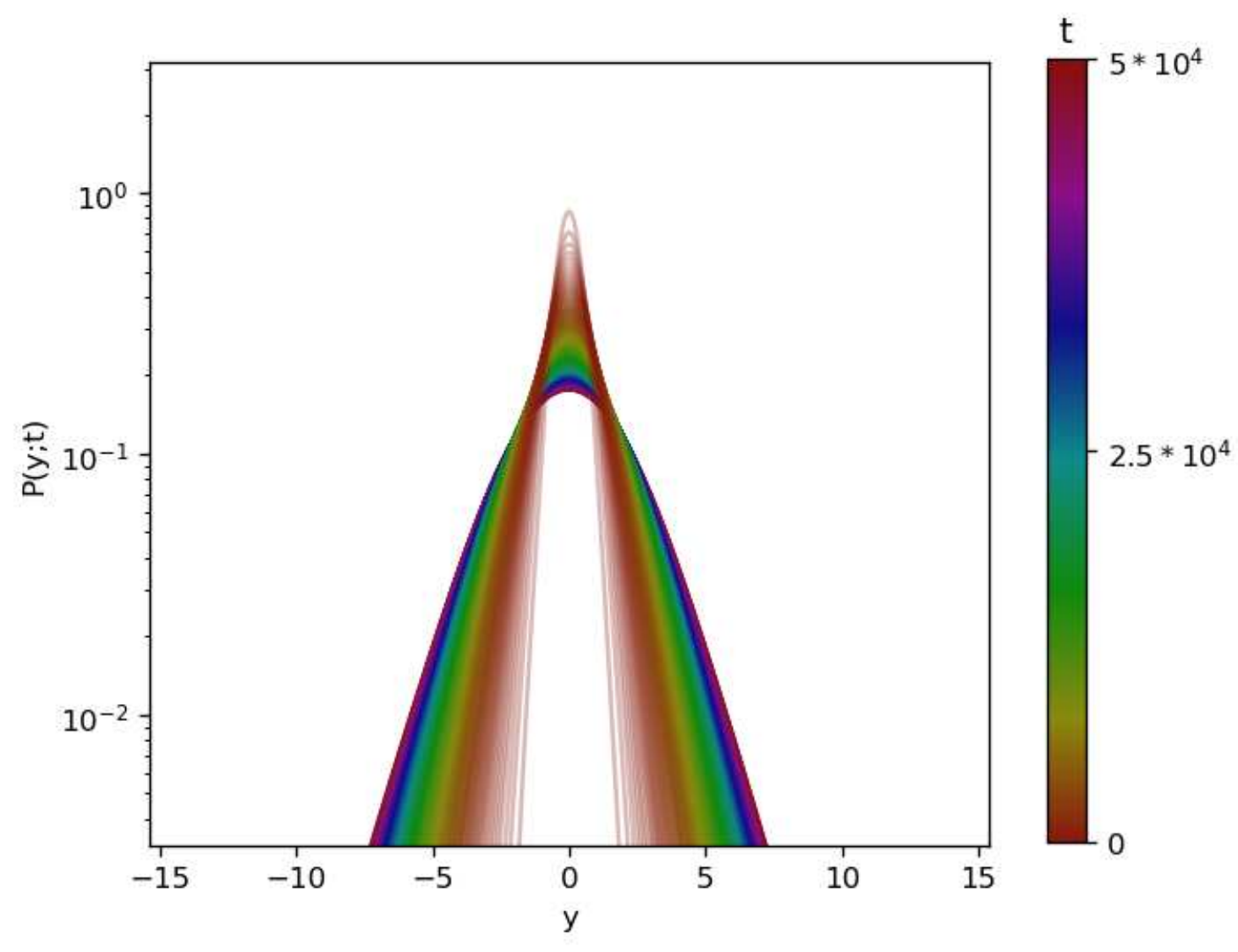}\\ (c) \includegraphics[width=0.45\textwidth]{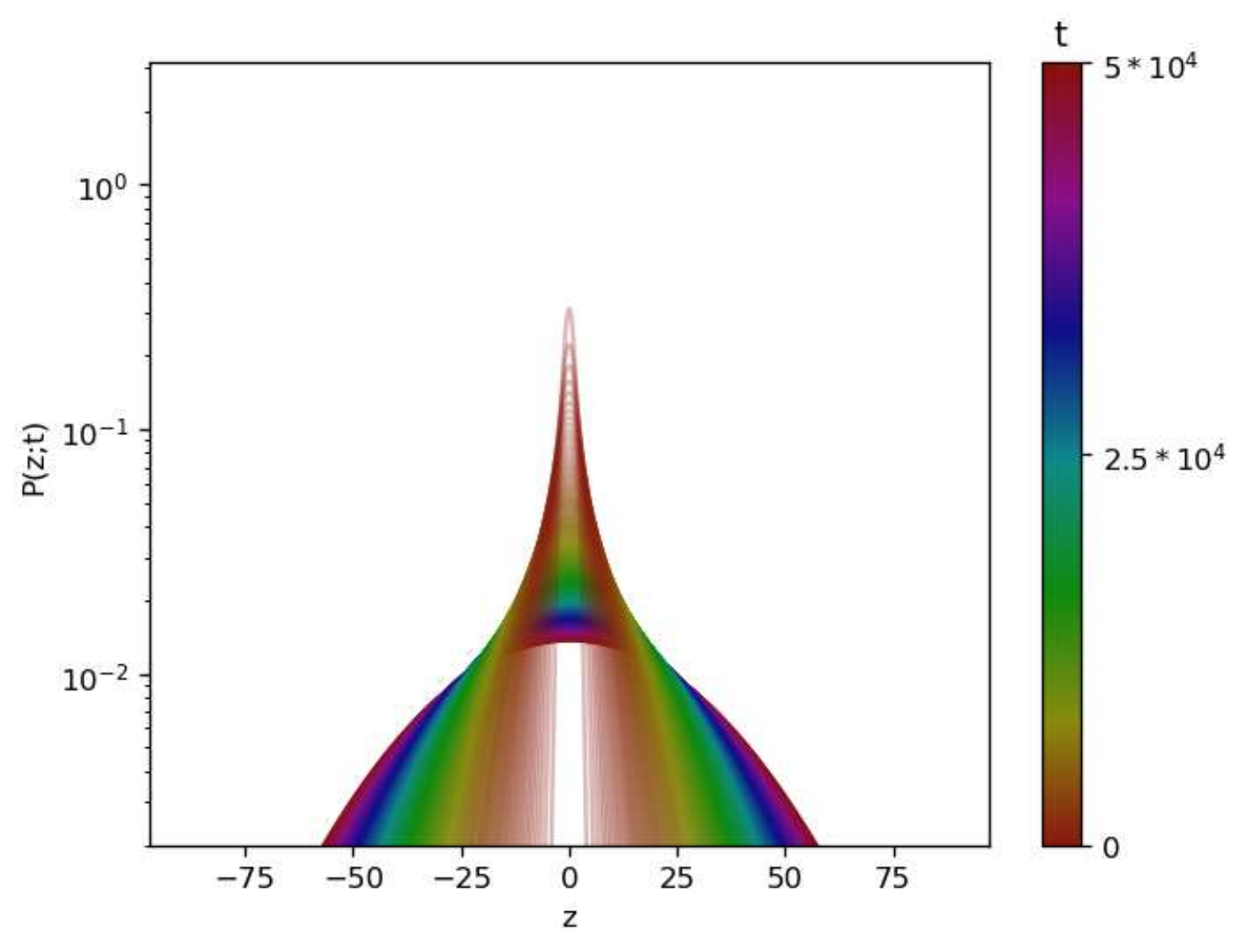}
\caption{Non-normalized PDF along (a) $x$ direction, (b) $y$ direction, and (c) $z$ direction, for $r=0$ (no resetting).}\label{PDF-globalX0}}
\end{figure}

\begin{figure}
\centering{(a) \includegraphics[width=0.45\textwidth]{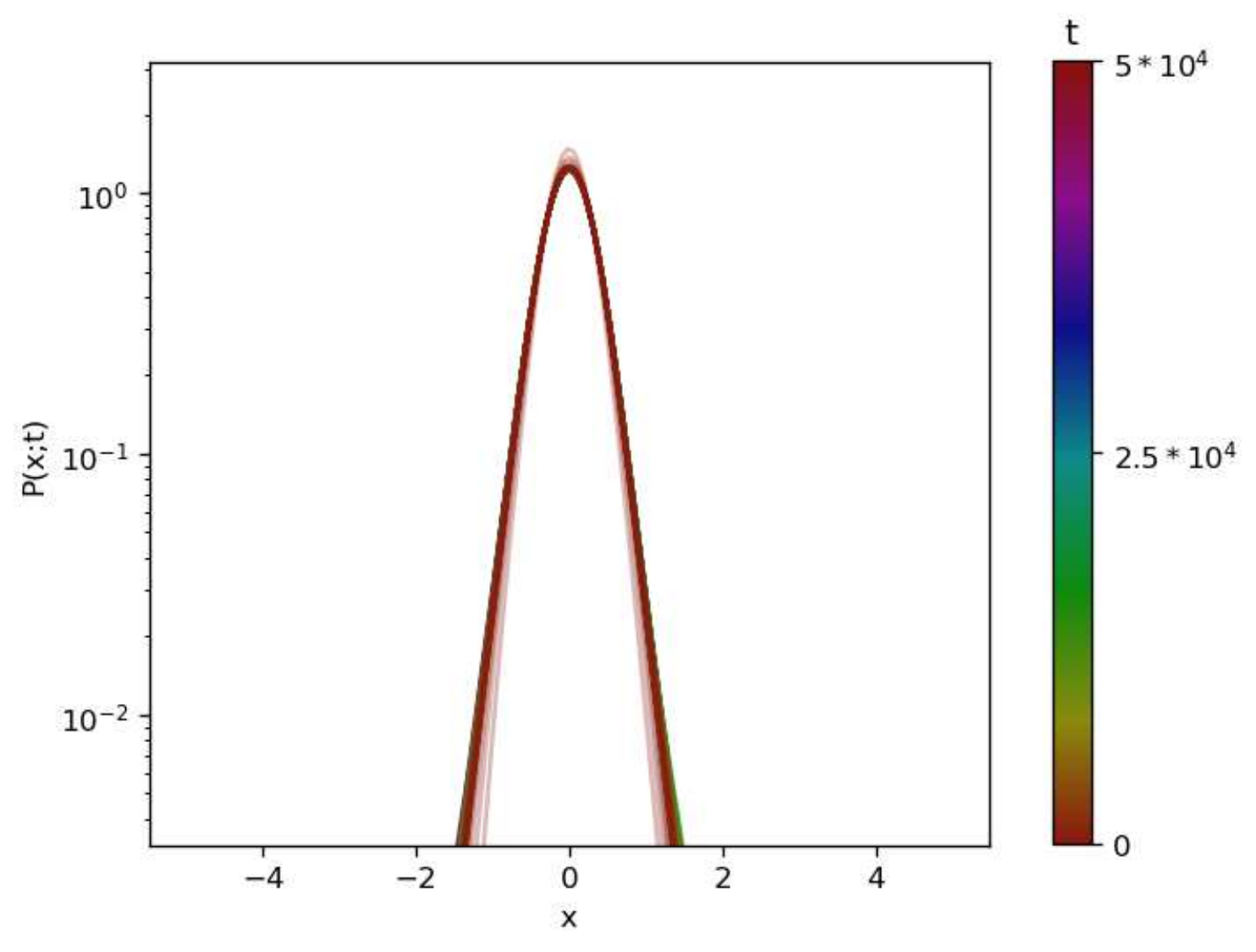} (b) \includegraphics[width=0.45\textwidth]{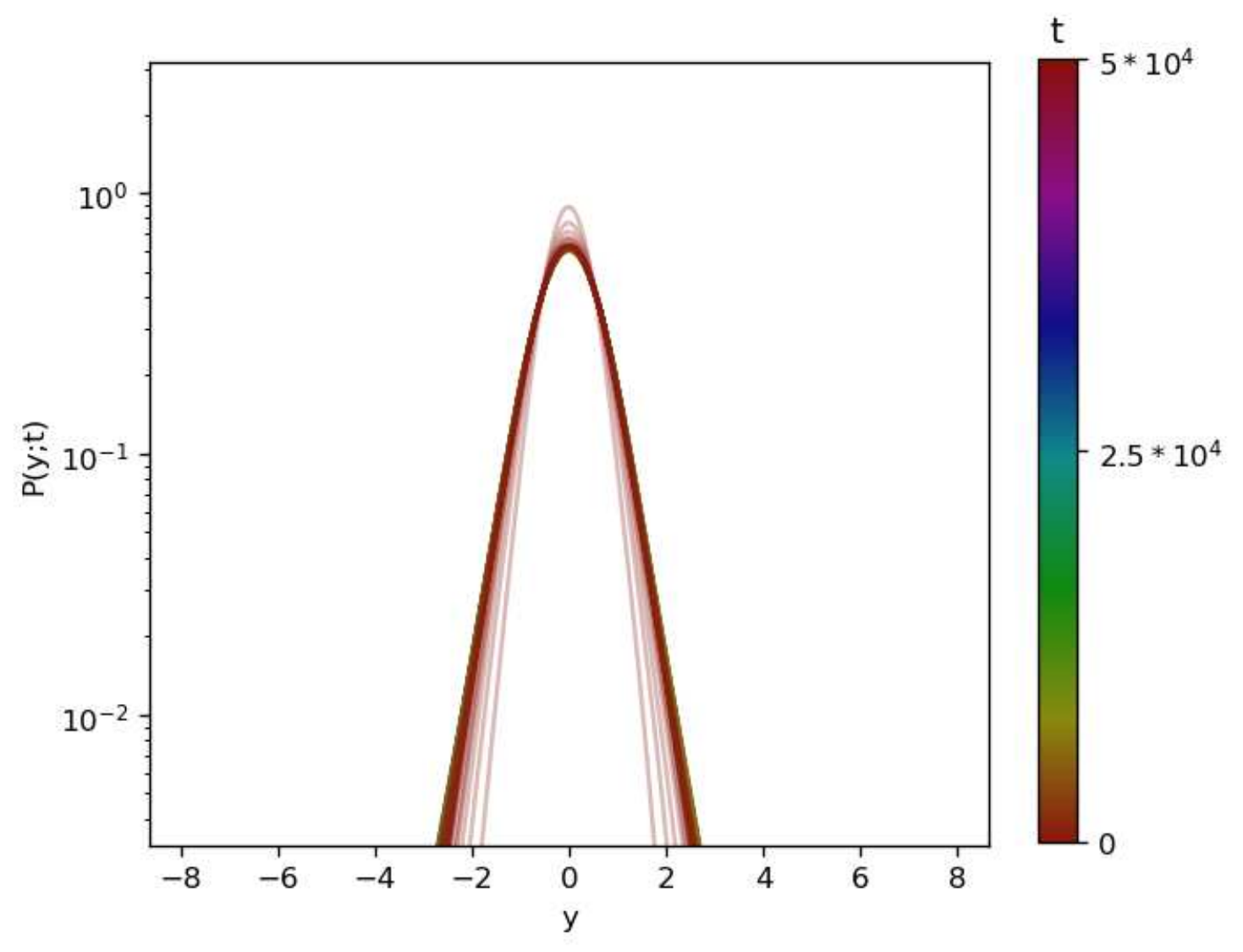}\\ (c) \includegraphics[width=0.45\textwidth]{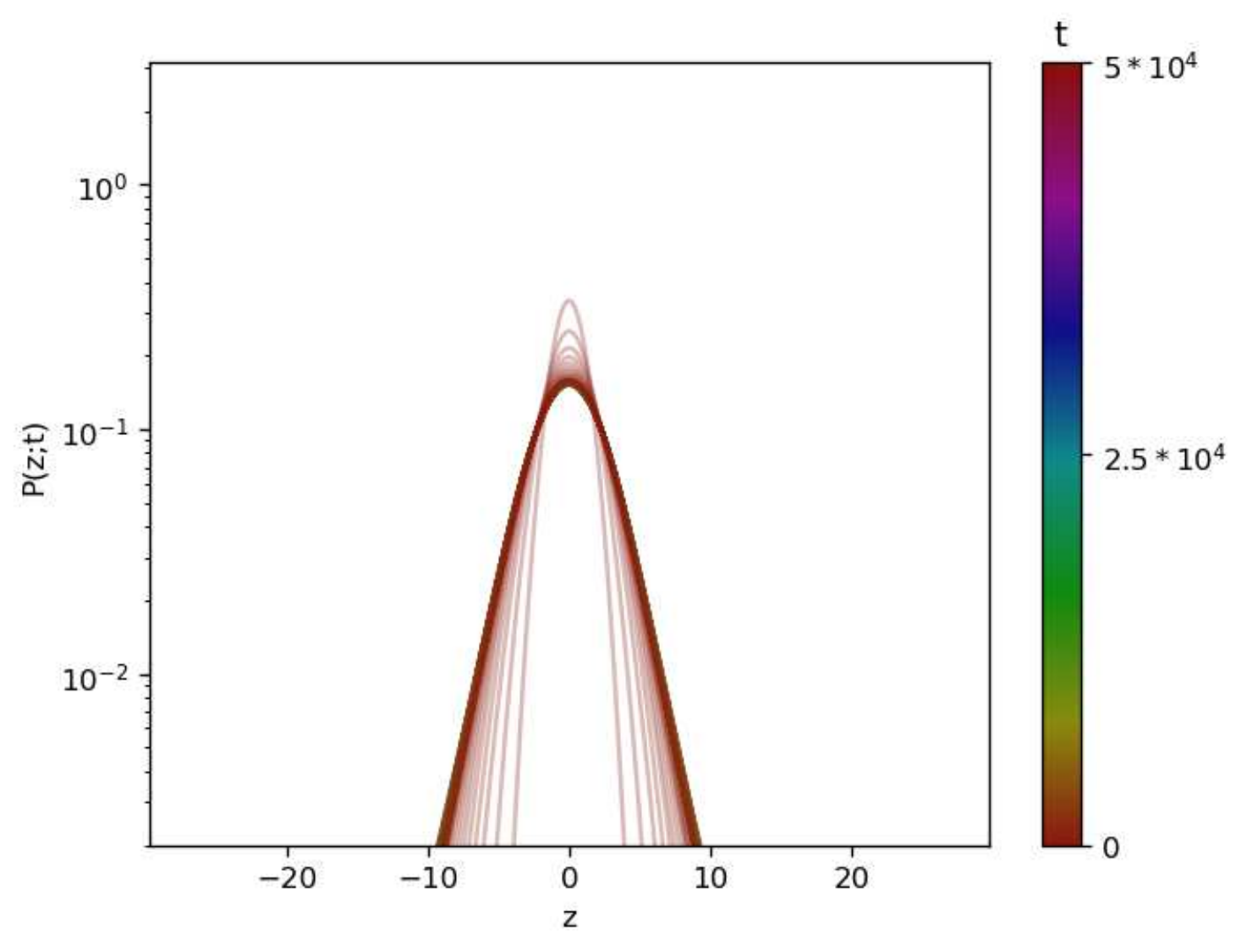}
\caption{Non-normalized PDF along (a) $x$ direction, (b) $y$ direction, and (c) $z$ direction, for global resetting  with rate $r=0.002$.}\label{PDF-globalX}}
\end{figure}

\subsection{Solutions for diffusion with reseting to the backbone}\label{app2}

Previously we have studied a resetting mechanism which takes the particle to a particular position of the comb, the coordinate $(x_0,0,0)$. Here, we proceed to analyse a slightly different resetting procedure, consisting on taking the particle to the backbone. This is, if the particle is at any position $(x,y,z)$, the resetting consists on taking it to the position $(x,0,0)$ of the state space. In this case, the Fokker-Planck equation reads
\begin{equation}
\begin{split}
\frac{\partial }{\partial t}P(x,y,z,t|x_0,0,0) &= \mathcal{D}_x\delta(y)\delta(z)\frac{\partial^2}{\partial x^2}P(x,y,z,t|x_0,0,0)+\mathcal{D}_y\delta(z)\frac{\partial^2}{\partial y^2}P(x,y,z,t|x_0,0,0)+\mathcal{D}_z\frac{\partial^2}{\partial z^2}P(x,y,z,t|x_0,0,0)\\
&-rP(x,y,z,t|x_0,0,0)+r\delta(y)\delta(z)\int_{-\infty}^\infty dy'\int_{-\infty}^{\infty}dz'P(x,y',z',t|x_0,0,0).
\end{split}
\label{EqFokkerPlanckResBackbone1}
\end{equation}
Here, instead of appearing at the particular position $(x_0,0,0)$ as stated by $\delta(x-x_0)\delta(y)\delta(z)$, the particle appears at $y=0,\ z=0$ but the $x$ position is not modified and, therefore, it can be mathematically written as the marginal distribution (double integral term in Eq.~\eqref{EqFokkerPlanckResBackbone1}). In order to solve the equation for the propagator $P(x,y,z,t|x_0,0,0)$, we apply the Fourier-Laplace transform as done in the previous case to obtain
\begin{equation}
\begin{split}
s\hat{P}(k_x,k_y,k_z,s|x_0,0,0)-e^{ik_x x_0} &= -\mathcal{D}_xk_x^2\hat{P}(k_x,y=0,z=0,s|x_0,0,0)-\mathcal{D}_yk_y^2\hat{P}(k_x,k_y,z=0,s|x_0,0,0)\\
&-\mathcal{D}_zk_z^2\hat{P}(k_x,k_y,k_z,s|x_0,0,0)-r\hat{P}(k_x,k_y,k_z,s|x_0,0,0)\\&+r\hat{P}(k_x,k_y=0,k_z=0,s|x_0,0,0).
\end{split}
\label{EqFokkerPlanckResBackbone1FL}
\end{equation}
Now, isolating the propagator one gets
\begin{equation}
\begin{split}
\hat{P}(k_x,k_y,k_z,s|x_0,0,0) &=\frac{e^{ik_x x_0}}{s+r+\mathcal{D}_zk_z^2}-\frac{ \mathcal{D}_xk_x^2\hat{P}(k_x,y=0,z=0,s|x_0,0,0)}{s+r+\mathcal{D}_zk_z^2}-\frac{\mathcal{D}_yk_y^2\hat{P}(k_x,k_y,z=0,s|x_0,0,0)}{s+r+\mathcal{D}_zk_z^2}\\&+\frac{r\hat{P}(k_x,k_y=0,k_z=0,s|x_0,0,0)}{s+r+\mathcal{D}_zk_z^2}.
\end{split}
\label{EqPropBackboneFL}
\end{equation}
Let us proceed to find the three unknown terms on the right hand side. Taking $k_y=k_z=0$ one can isolate
\begin{equation}
\hat{P}(k_x,k_y=0,k_z=0,s|x_0,0,0)=\frac{e^{ik_x x_0}}{s}-\frac{\mathcal{D}_xk_x^2\hat{P}(k_x,y=0,z=0,s|x_0,0,0)}{s},
\label{EqPropBackboneFL1}
\end{equation}
with which the equation for the propagator reads
\begin{equation}
\begin{split}
\hat{P}(k_x,k_y,k_z,s|x_0,0,0) &=\frac{s+r}{s+r+\mathcal{D}_zk_z^2}\frac{e^{ik_x x_0}}{s}-\frac{ s+r}{s+r+\mathcal{D}_zk_z^2}\frac{\mathcal{D}_xk_x^2}{s}\hat{P}(k_x,y=0,z=0,s|x_0,0,0)\\&-\frac{\mathcal{D}_yk_y^2\hat{P}(k_x,k_y,z=0,s|x_0,0,0)}{s+r+\mathcal{D}_zk_z^2}.
\end{split}
\label{EqPropBackboneFL2}
\end{equation}
The other two terms can be obtained by taking the inverse Fourier transform of $k_y$ and $k_z$ and evaluating at $y=0$ and $z=0$ respectively. Starting by the last, one gets that
\begin{equation}
\begin{split}
\hat{P}(k_x,k_y,z=0,s|x_0,0,0) &=\frac{s+r}{2\sqrt{D_z(s+r)}+\mathcal{D}_yk_y^2}\frac{1}{s}\left(e^{ik_x x_0}-\mathcal{D}_xk_x^2\hat{P}(k_x,y=0,z=0,s|x_0,0,0)\right).
\end{split}
\label{EqPropBackboneFL3}
\end{equation}
Proceeding similarly for the $y$ variable one gets
\begin{equation}
\begin{split}
\hat{P}(k_x,y=0,z=0,s|x_0,0,0) &=\frac{e^{ik_x x_0}}{2\frac{s}{s+r}\sqrt{2D_y\sqrt{D_z(s+r)}}+\mathcal{D}_xk_x^2}.
\end{split}
\label{EqPropBackboneFL4}
\end{equation}
Putting all the elements into Eq.~\eqref{EqPropBackboneFL2} and simplifying some terms one gets the following explicit expression for the propagator:
\begin{equation}
\begin{split}
\hat{P}(k_x,k_y,k_z,s|x_0,0,0) &=\frac{e^{ik_x x_0}4\sqrt{2\mathcal{D}_y\sqrt{\mathcal{D}_z(s+r)}}\sqrt{\mathcal{D}_z(s+r)}}{\left( s+r+\mathcal{D}_zk_z^2\right)\left( 2\sqrt{\mathcal{D}_z(s+r)}+\mathcal{D}_yk_y^2\right)\left( 2\frac{s}{s+r}\sqrt{2\mathcal{D}_y\sqrt{\mathcal{D}_z(s+r)}}+\mathcal{D}_xk_x^2\right)}.
\end{split}
\label{EqPropBackboneFL2_0}
\end{equation}

Finally, from these expressions one can get the corresponding MSD,
\begin{equation}
\langle x^2(t)\rangle=x_0^2+\frac{\mathcal{D}_x}{\sqrt{2\mathcal{D}_y\sqrt{\mathcal{D}_z}}}\,\mathcal{L}^{-1}\left[\frac{(s+r)^{3/4}}{s^2}\right]=x_0^2+\frac{\mathcal{D}_x}{\sqrt{2\mathcal{D}_y\sqrt{\mathcal{D}_z}}}\,t^{1/4}\,E_{1,5/4}^{-3/4}(-r\,t),
\label{EqMSDxaxis2}
\end{equation}
\begin{equation}
\langle y^2(t)\rangle=\frac{\mathcal{D}_y}{\sqrt{\mathcal{D}_z}}\,\mathcal{L}^{-1}\left[\frac{1}{s(s+r)^{1/2}}\right]=\frac{\mathcal{D}_y}{\sqrt{\mathcal{D}_z}}\frac{\mathrm{erf}(\sqrt{rt})}{\sqrt{r}}
\end{equation}
\begin{equation}
\langle z^2(t)\rangle=2\,\mathcal{D}_z\,\mathcal{L}^{-1}\left[\frac{1}{s(s+r)}\right]=2\,\mathcal{D}_z\,\frac{1-e^{-rt}}{r}.
\end{equation}

In Fig.~\ref{PDF-backX}, 
we give graphical representations of the PDFs of the process under resetting to the backbone.

\begin{figure}
\centering{(a) \includegraphics[width=0.45\textwidth]{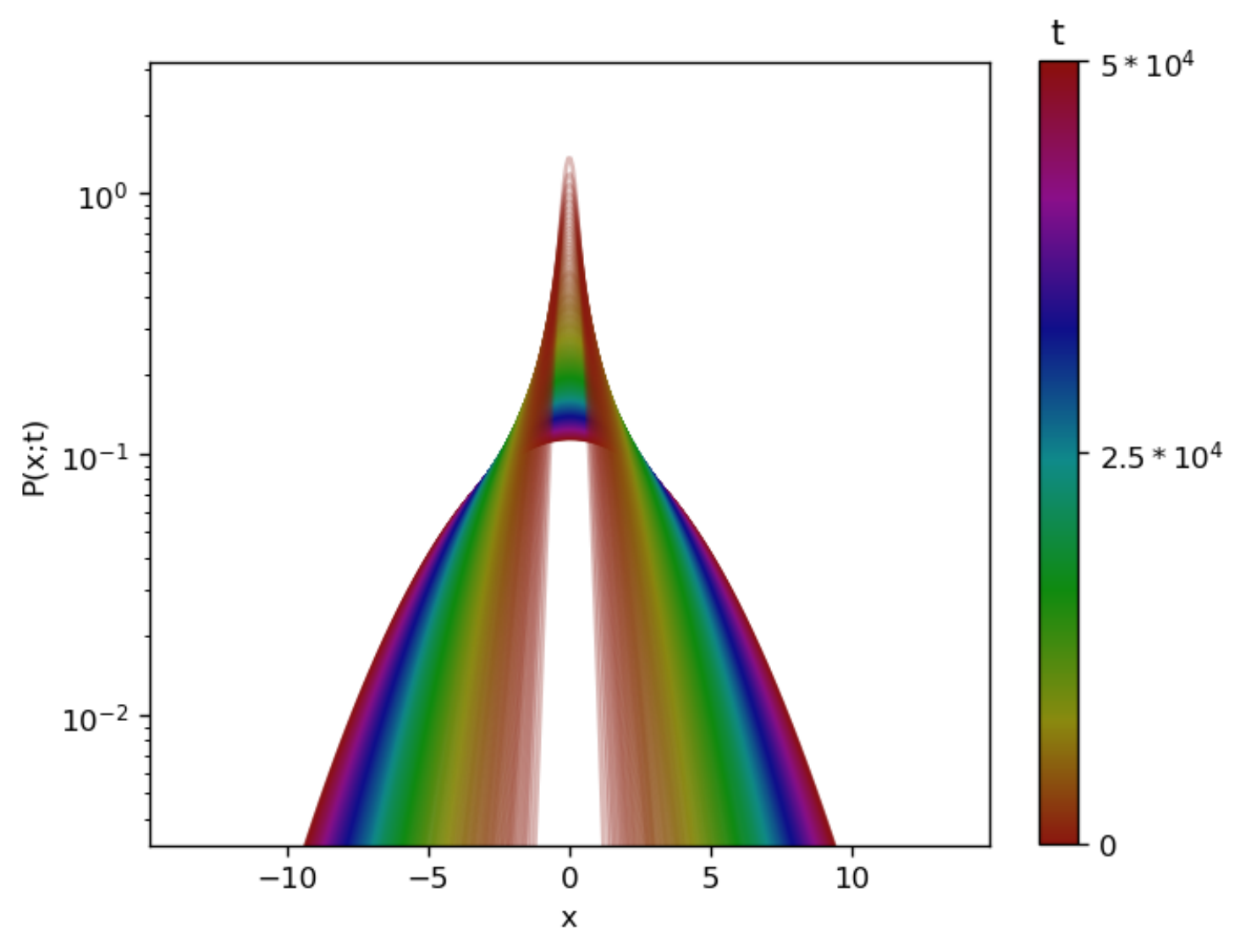} (b) \includegraphics[width=0.45\textwidth]{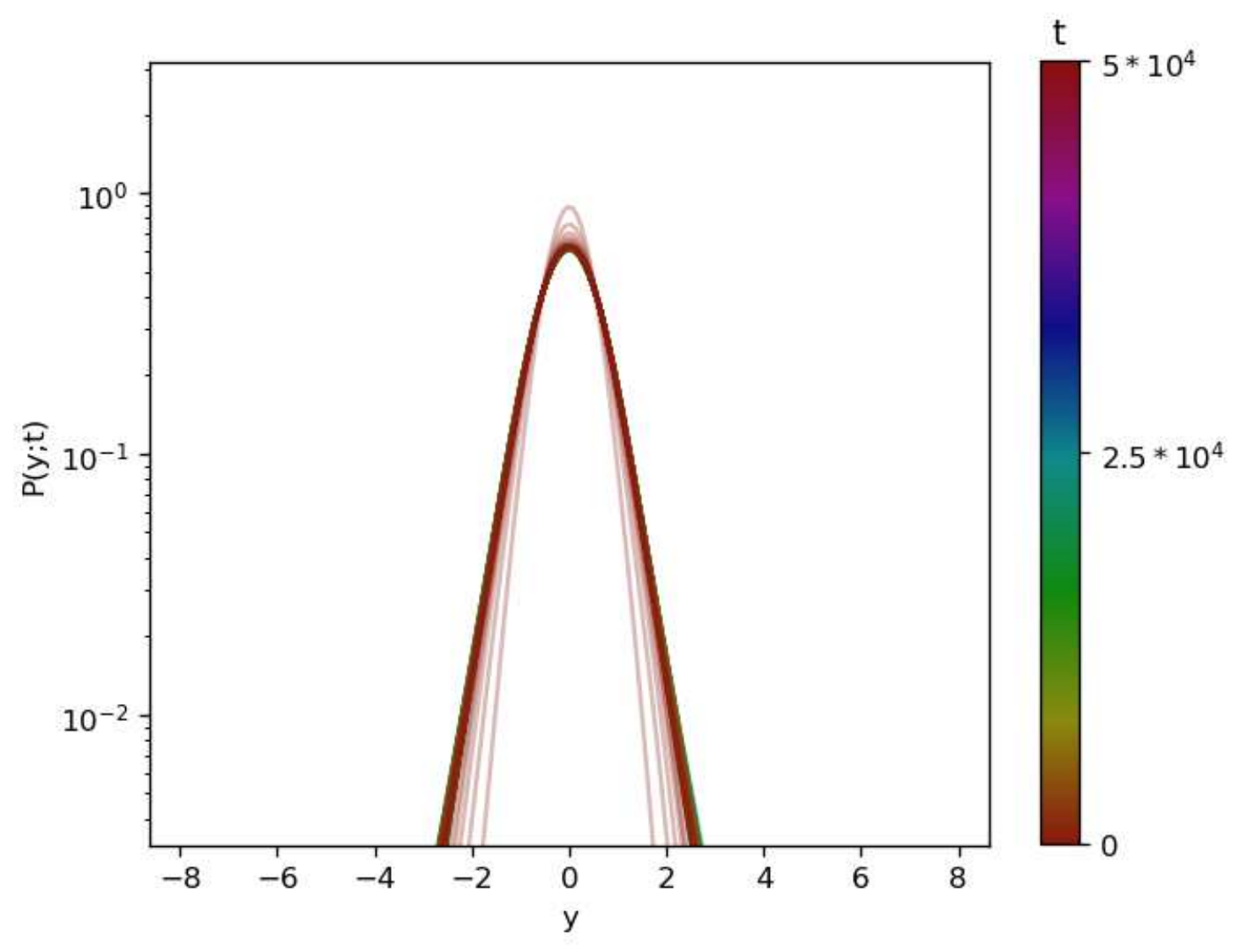}\\ (c) \includegraphics[width=0.45\textwidth]{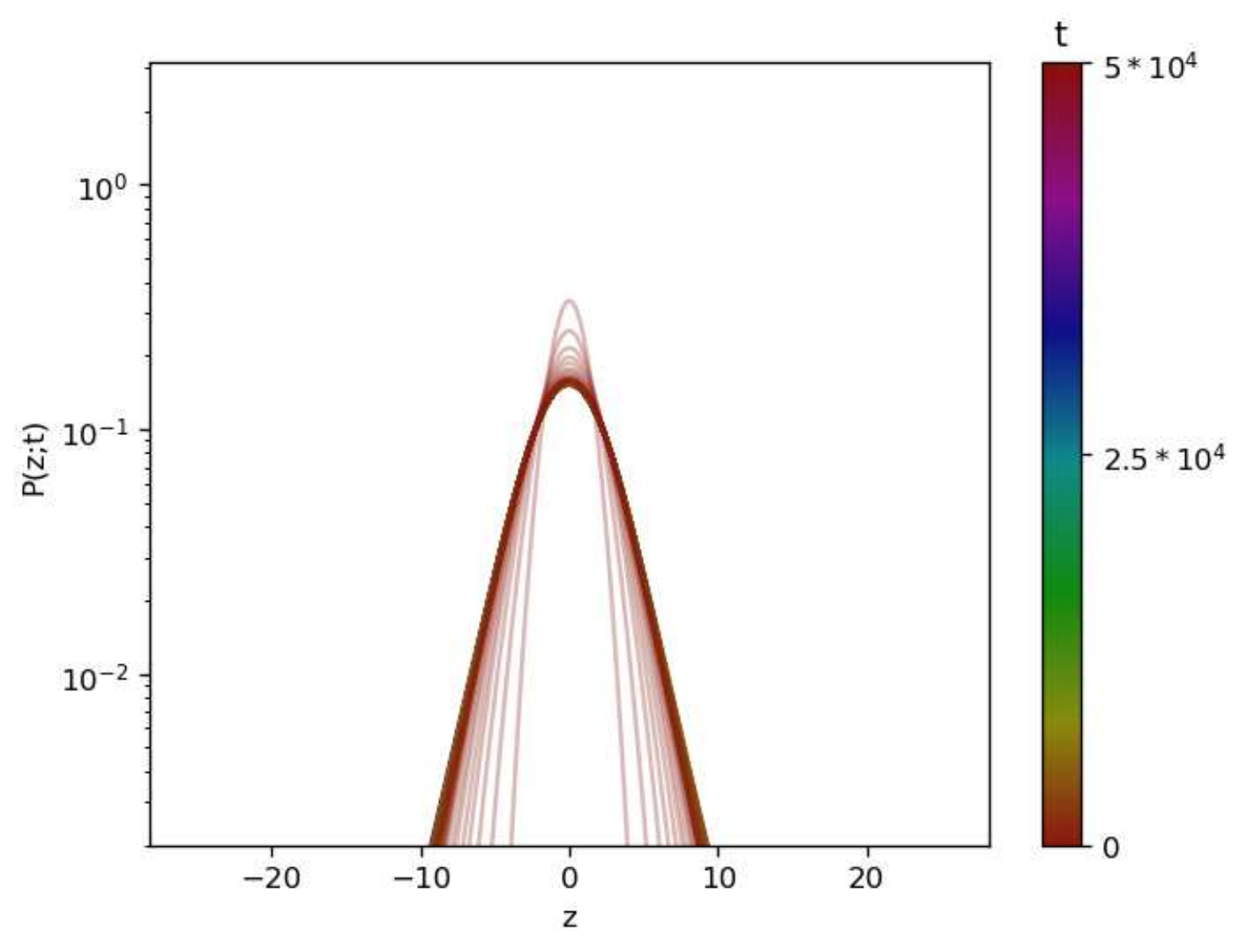}
\caption{Non-normalized PDF along (a) $x$ direction, (b) $y$ direction, and (c) $z$ direction, for resetting to the backbone with rate $r=0.002$.}\label{PDF-backX}}
\end{figure}

\subsection{Solutions for diffusion with reseting to the main fingers}\label{app3}

Finally, we study the dynamics of the system when, instead of taking the particle to a fixed position in the $x$ axis or to the backbone, the resetting only applies to the $z$ axis, taking the particle to the corresponding main finger. In this case, the governing equation reads
\begin{equation}
\begin{split}
\frac{\partial }{\partial t}P(x,y,z,t|x_0,0,0) &= \mathcal{D}_x\delta(y)\delta(z)\frac{\partial^2}{\partial x^2}P(x,y,z,t|x_0,0,0)+\mathcal{D}_y\delta(z)\frac{\partial^2}{\partial y^2}P(x,y,z,t|x_0,0,0)+\mathcal{D}_z\frac{\partial^2}{\partial z^2}P(x,y,z,t|x_0,0,0)\\
&-rP(x,y,z,t|x_0,0,0)+r\delta(z)\int_{-\infty}^{\infty}dz'P(x,y',z',t|x_0,0,0),
\end{split}
\label{EqFokkerPlanckResFingers1}
\end{equation}
where the last term is now the marginal distribution in the variables $x$ and $y$ instead of only the $x$ variable. The corresponding equation in the Fourier-Laplace space reads
\begin{equation}
\begin{split}
s\hat{P}(k_x,k_y,k_z,s|x_0,0,0)-e^{ik_x x_0} &= -\mathcal{D}_xk_x^2\hat{P}(k_x,y=0,z=0,s|x_0,0,0)-\mathcal{D}_yk_y^2\hat{P}(k_x,k_y,z=0,s|x_0,0,0)\\
&-\mathcal{D}_zk_z^2\hat{P}(k_x,k_y,k_z,s|x_0,0,0)-r\hat{P}(k_x,k_y,k_z,s|x_0,0,0)\\&+r\hat{P}(k_x,k_y,k_z=0,s|x_0,0,0),
\end{split}
\label{EqFokkerPlanckResFingers1FL}
\end{equation}
from which the propagator can be isolated to give
\begin{equation}
\begin{split}
\hat{P}(k_x,k_y,k_z,s|x_0,0,0) &=\frac{e^{ik_x x_0}}{s+r+\mathcal{D}_zk_z^2}-\frac{ \mathcal{D}_xk_x^2\hat{P}(k_x,y=0,z=0,s|x_0,0,0)}{s+r+\mathcal{D}_zk_z^2}-\frac{\mathcal{D}_yk_y^2\hat{P}(k_x,k_y,z=0,s|x_0,0,0)}{s+r+\mathcal{D}_zk_z^2}\\&+\frac{r\hat{P}(k_x,k_y,k_z=0,s|x_0,0,0)}{s+r+\mathcal{D}_zk_z^2}.
\end{split}
\label{EqPropFingersFL}
\end{equation}
and, by taking $k_z=0$, we can isolate the following expression for the marginal PDF:

\begin{equation}
\begin{split}
\hat{P}(k_x,k_y,k_z=0,s|x_0,0,0)&=\frac{e^{ik_x x_0}}{s}-\frac{\mathcal{D}_xk_x^2\hat{P}(k_x,y=0,z=0,s|x_0,0,0)}{s}
-\frac{\mathcal{D}_yk_y^2\hat{P}(k_x,k_y,z=0,s|x_0,0,0)}{s}.
\end{split}
\label{EqPropFingersFL1}
\end{equation}
Introducing this expression to Eq.~\eqref{EqPropFingersFL} we get
\begin{equation}
\begin{split}
\hat{P}(k_x,k_y,k_z,s|x_0,0,0) &=\frac{s+r}{s(s+r+\mathcal{D}_z\,k_z^2)}\left(e^{ik_x x_0}-\mathcal{D}_x\,k_x^2\hat{P}(k_x,y=0,z=0,s|x_0,0,0)-\mathcal{D}_y\,k_y^2\hat{P}(k_x,k_y,z=0,s|x_0,0,0)\right).
\end{split}
\label{EqPropFingersFL2}
\end{equation}
As done for the previous cases, the two unknown terms remaining on the right hand side can be found by applying the inverse Fourier transform with respect to the $y$ and $z$ variables to Eq.~\eqref{EqPropFingersFL2} and evaluating at $y=z=0$. Starting by the last, one finds that
\begin{equation}
\begin{split}
\hat{P}(k_x,k_y,z=0,s|x_0,0,0) &=\frac{1}{2\frac{s}{s+r}\sqrt{\mathcal{D}_z(s+r)}+\mathcal{D}_yk_y^2}\left(e^{ik_x x_0}-\mathcal{D}_x\,k_x^2\hat{P}(k_x,y=0,z=0,s|x_0,0,0)\right)
\end{split}
\label{EqPropFingersFL3}
\end{equation}
and proceeding equally for the $y$ variable:
\begin{equation}
\begin{split}
\hat{P}(k_x,y=0,z=0,s|x_0,0,0) &=\frac{e^{ik_x x_0}}{2\sqrt{2\mathcal{D}_y\frac{s}{s+r}\sqrt{\mathcal{D}_z(s+r)}}+\mathcal{D}_x\,k_x^2}.
\end{split}
\label{EqPropFingersFL3}
\end{equation}
Putting all the elements together one finds a formal expression for the propagator:
\begin{equation}
\begin{split}
\hat{P}(k_x,k_y,k_z,s|x_0,0,0) &=\frac{4e^{ik_x x_0}\sqrt{\mathcal{D}_z(s+r)}\sqrt{2\mathcal{D}_y\frac{s}{s+r}\sqrt{\mathcal{D}_z(s+r)}}}{(s+r+\mathcal{D}_z\,k_z^2)\left(2\frac{s}{s+r}\sqrt{\mathcal{D}_z(s+r)}+\mathcal{D}_y\,k_y^2 \right)\left( 2\sqrt{2\mathcal{D}_y\frac{s}{s+r}\sqrt{\mathcal{D}_z(s+r)}}+\mathcal{D}_x\,k_x^2\right)}.
\end{split}
\label{EqPropFingersFL2_0}
\end{equation}

Finally, the corresponding MSD in the Laplace space read
\begin{equation}
\langle x^2(t)\rangle=x_0^2+\frac{\mathcal{D}_x}{\sqrt{2\mathcal{D}_y\sqrt{\mathcal{D}_z}}}\,\mathcal{L}^{-1}\left[\frac{(s+r)^{1/4}}{s^{3/2}}\right]=x_{0}^{2}+\frac{\mathcal{D}_x}{\sqrt{2\mathcal{D}_y\sqrt{\mathcal{D}_z}}}\,t^{1/4}\,E_{1,5/4}^{-1/4}(-r\,t),
\label{EqMSDxaxis2}
\end{equation}
\begin{equation}
\langle y^2(t)\rangle=\frac{\mathcal{D}_y}{\sqrt{\mathcal{D}_z}}\,\mathcal{L}^{-1}\left[\frac{(s+r)^{1/2}}{s^2}\right]=\frac{\mathcal{D}_y}{\sqrt{\mathcal{D}_z}}\,t^{1/2}\,E_{1,3/2}^{-1/2}(-r\,t)=\frac{\mathcal{D}_y}{\sqrt{\mathcal{D}_z}}\left[e^{-rt}\frac{t^{1/2}}{\Gamma(1/2)}+\frac{2rt+1}{2\sqrt{r}}\,\mathrm{erf}(\sqrt{rt})\right],
\end{equation}
\begin{equation}
\langle z^2(t)\rangle=2\,\mathcal{D}_z\,\mathcal{L}^{-1}\left[\frac{1}{s(s+r)}\right]=\frac{2\,\mathcal{D}_{z}}{r}\left[1-e^{-rt}\right].
\end{equation}

In Fig.~\ref{PDF-finX}, 
we give graphical representations of the PDFs of the process under resetting to the backbone.

\begin{figure}
\centering{(a) \includegraphics[width=0.45\textwidth]{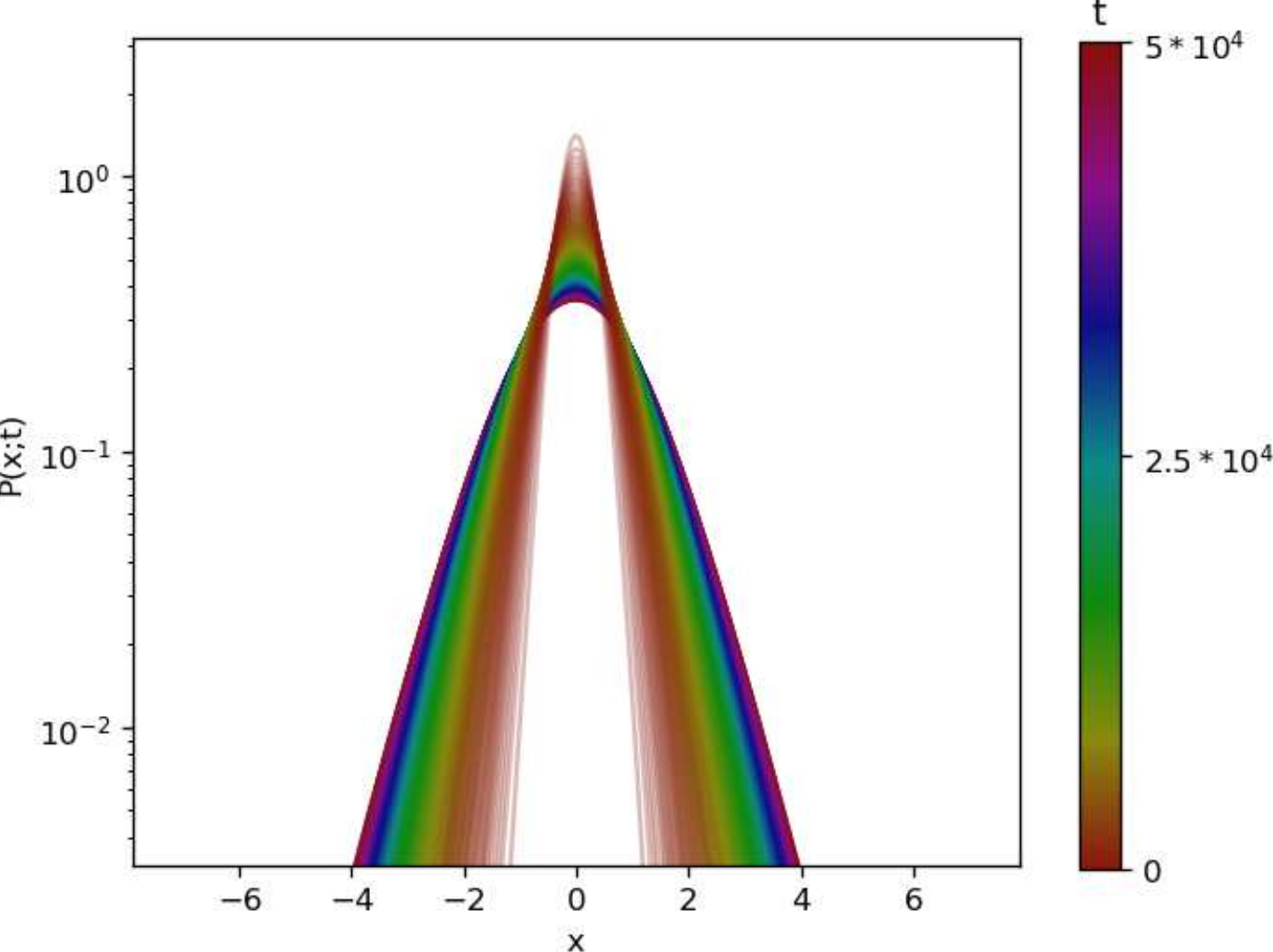} (b) \includegraphics[width=0.45\textwidth]{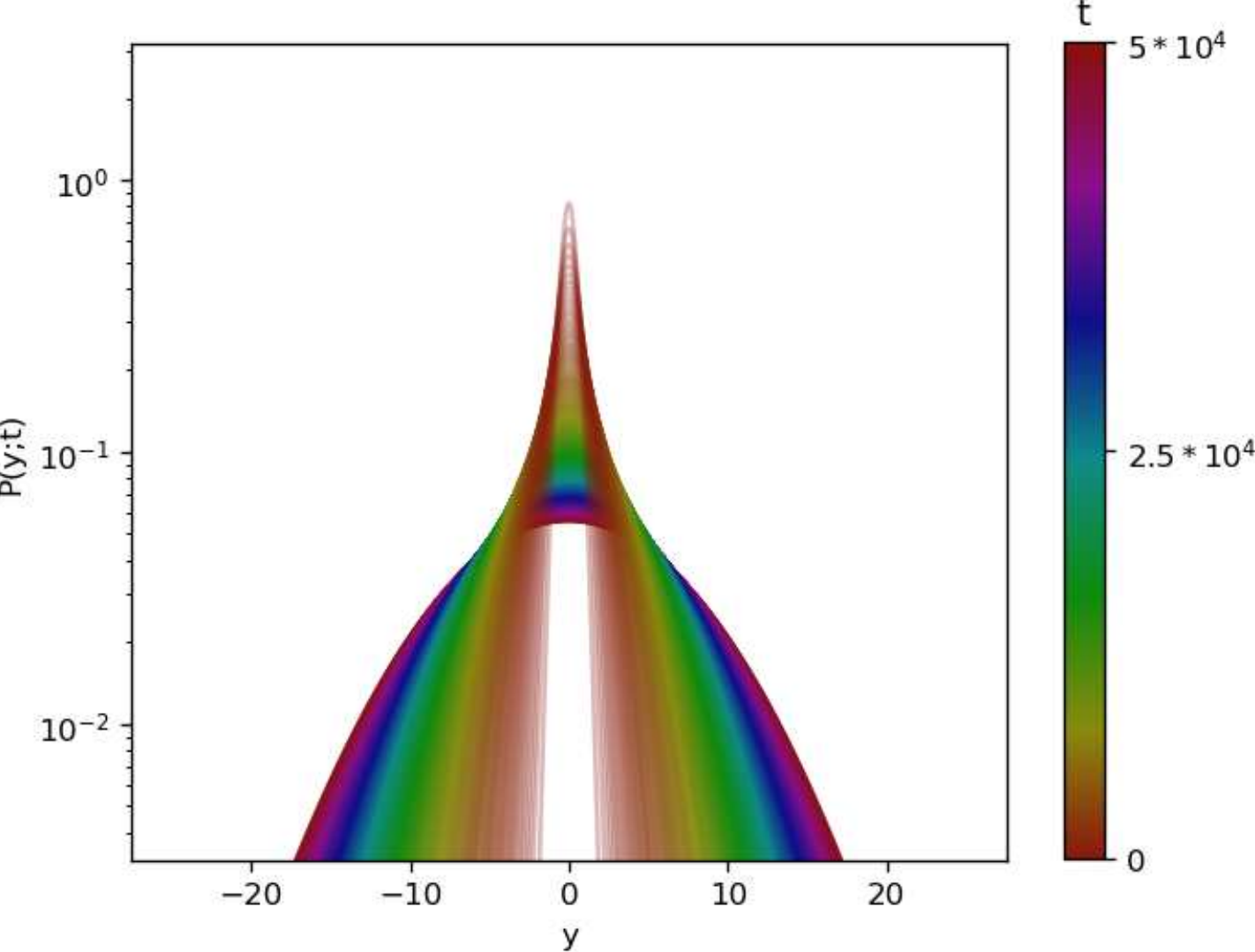}\\ (c) \includegraphics[width=0.45\textwidth]{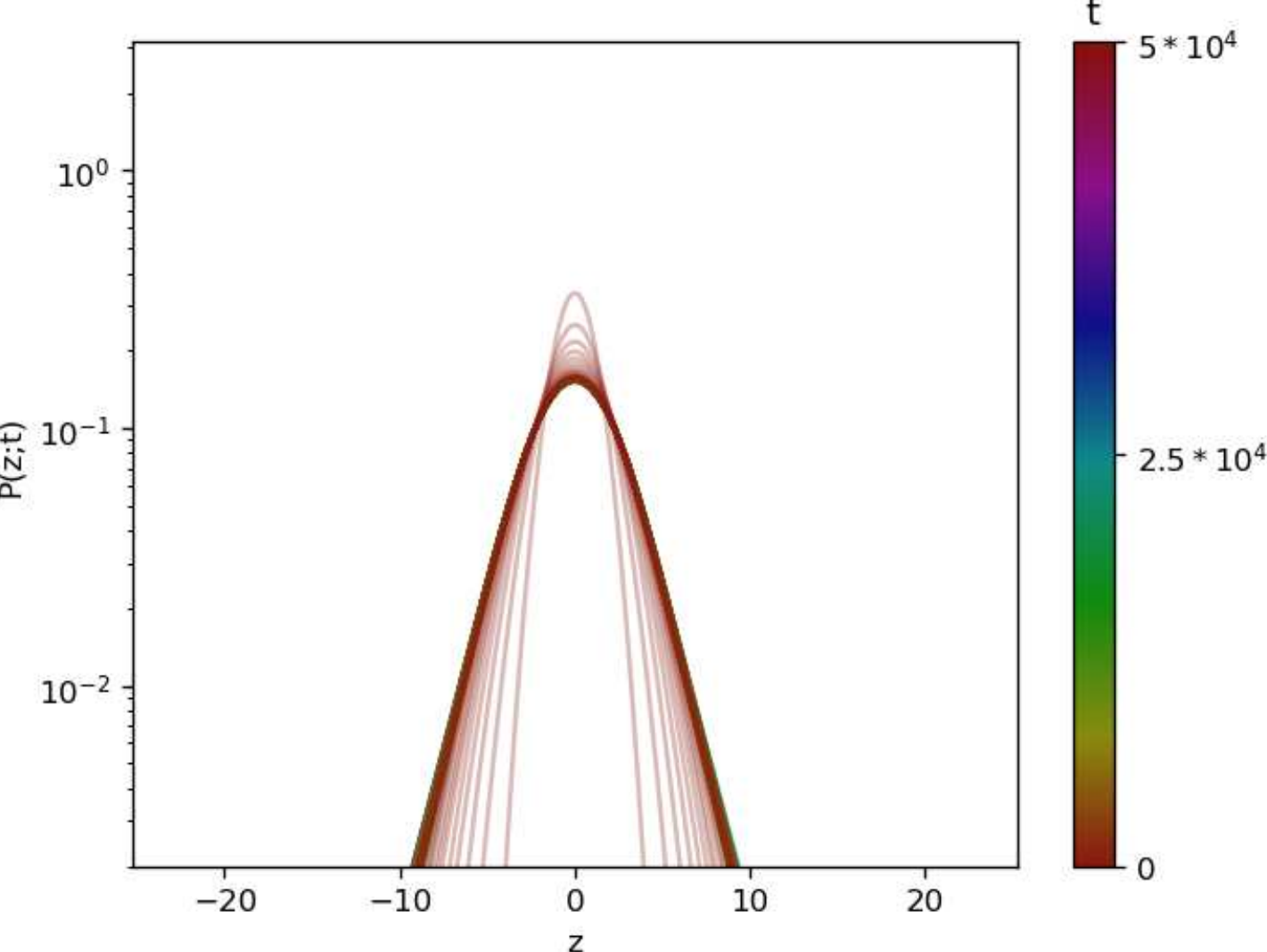}
\caption{Non-normalized PDF along (a) $x$ direction, (b) $y$ direction, and (c) $z$ direction, for resetting to the fingers with rate $r=0.002$.}\label{PDF-finX}}
\end{figure}

\end{widetext}

\end{document}